\documentclass[amsmath,twocolumn]{aastex702}

\usepackage{amsmath}
\usepackage{xcolor}
\usepackage{soul}
\usepackage{upgreek}
\usepackage{multirow}
\usepackage{comment}
\shorttitle{X-ray Analysis of MACSJ1752 \& ZWCL1856}

\shortauthors{HyeongHan et al.}
\graphicspath{{./}{figures/}}

\begin{document}

\title{Exemplary Merging Clusters II: XMM-{\it Newton} and {\it Chandra} X-ray Observations of MACS~J1752.0+4440 and ZwCl~1856.8+6616}

\correspondingauthor{K. HyeongHan}

\author[0000-0002-2550-5545]{K. HyeongHan}
\affiliation{Department of Physics, Duke University, Durham, NC 27708, USA}
\email[show]{hyeonghan.kim@duke.edu}

\author[0000-0002-4462-0709]{K. Finner}
\affiliation{IPAC, California Institute of Technology, 1200 E California Blvd., Pasadena, CA 91125, USA}
\email{kfinner@ipac.caltech.edu}

\author[0000-0002-1566-5094]{W. Lee}
\affiliation{Max-Planck-Institut f{\"u}r Astronomie, K{\"o}nigstuhl 17, D-69117 Heidelberg, Germany}
\email{wonkilee@mpia.de}

\author[0000-0002-1616-5649]{E. M. Silich}
\affiliation{Cahill Center for Astronomy and Astrophysics, California Institute of Technology, Pasadena, CA 91125, USA}
\email{esilich@astro.caltech.edu}

\author[0000-0002-9478-1682]{W. Forman}
\affiliation{Center for Astrophysics $\vert$ Harvard \& Smithsonian, 60 Garden Street, Cambridge, MA 02138, USA}
\email{wforman@cfa.harvard.edu}

\author[0000-0003-2206-4243]{C. Jones}
\affiliation{Center for Astrophysics $\vert$ Harvard \& Smithsonian, 60 Garden Street, Cambridge, MA 02138, USA}
\email{cjones@cfa.harvard.edu}

\author[0000-0002-3132-8776]{A. T\"{u}mer}
\affiliation{Center for Space Science and Technology, University of Maryland, Baltimore County (UMBC), Baltimore, MD 21250, USA}
\affiliation{NASA/Goddard Space Flight Center, Greenbelt, MD 20771, USA}
\affiliation{Center for Research and Exploration in Space Science and Technology, NASA/GSFC (CRESST II), Greenbelt, MD 20771, USA}
\email{aysegult@umbc.edu}

\author[0000-0002-3984-4337]{S. Randall}
\affiliation{Center for Astrophysics $\vert$ Harvard \& Smithsonian, 60 Garden Street, Cambridge, MA 02138, USA}
\email{srandall@cfa.harvard.edu}

\author[0000-0003-3175-2347]{J. ZuHone}
\affiliation{Center for Astrophysics $\vert$ Harvard \& Smithsonian, 60 Garden Street, Cambridge, MA 02138, USA}
\email{john.zuhone@cfa.harvard.edu}

\author[0000-0002-5751-3697]{M. James Jee}
\affiliation{Department of Astronomy, Yonsei University, 50 Yonsei-ro, Seoul 03722, Korea}
\affiliation{Department of Physics, University of California, Davis, One Shields Avenue, Davis, CA 95616, USA}
\email{mkjee@yonsei.ac.kr}

\author[0000-0002-0813-5888]{D. Wittman}
\affiliation{Department of Physics and Astronomy, University of California, Davis, CA 95616, USA}
\email{dwittman@physics.ucdavis.edu}

\author[0000-0002-0587-1660]{R. J. van Weeren}
\affiliation{Leiden Observatory, Leiden University, PO Box 9513, 2300 RA Leiden, The Netherlands}
\email{rvweeren@strw.leidenuniv.nl}

\author{M. Troxel}
\affiliation{Department of Physics, Duke University, Durham, NC 27708, USA}
\email{michael.troxel@duke.edu}

\begin{abstract}

Using deep XMM-{\it Newton} and {\it Chandra} observations, we compare the double-radio-relic galaxy cluster mergers MACS~J1752.0+4440 and ZwCl~1856.8+6616. 
Both are 1:1 mass-ratio mergers with similar relic separations but differing nearly an order of magnitude in total mass. 
In MACS~J1752.0+4440, the surface-brightness discontinuity at the northeastern relic yields $\mathcal{M}_{\rho}=1.4\pm0.2$, while the temperature jump provides a weaker constraint, corresponding to $\mathcal{M}_{kT}\lesssim2.3$ at the $2\sigma$ level, lower than the radio-derived $\mathcal{M}_{\rm radio}=3.1\pm0.1$.
No discontinuity is detected at the southwestern relic. 
Thermodynamic maps and axial profiles reveal surviving relatively low-entropy cores embedded in Mpc-scale stripped-gas structures. 
The nearly uniform X-ray tail brightness and a southwestern plateau coincident with a third lensing mass peak favor a non-zero impact parameter and complex assembly history. 
In ZwCl~1856.8+6616, we do not detect a clear surface brightness discontinuity at either relic. 
The northern X-ray peak leads its brightest cluster galaxy (BCG) and lensing peak, while the hotter southern X-ray peak is laterally offset from its BCG and lensing peak. 
The asymmetric intracluster emission favors a phase near or after first apocenter, when displaced gas can develop a slingshot-like morphology. 
We present two TNG-Cluster analogs in which a minor merger followed by an off-axis major collision produces extended X-ray tails resembling MACS~J1752.0+4440 and later leads to gas structures resembling ZwCl~1856.8+6616. 
X-ray morphology independently confirms the merger phases inferred from relic-based chronometry, placing MACS~J1752.0+4440 before and ZwCl~1856.8+6616 after first apocenter, and demonstrating that similar relic and lensing configurations can correspond to distinct dynamical states.
\end{abstract}


\section{Introduction}
Major galaxy cluster mergers are among the most energetic events in the universe, releasing kinetic energies on the order of $10^{64}$ ergs \citep{2002sarazin, 2007markevitch}. 
As merging clusters accelerate toward each other and approach first pericenter passage, their relative velocity can become supersonic.
The resulting collision is governed by the different physical properties of the cluster components.
The hot intracluster medium \citep[ICM; e.g.,][]{Forman1982, Jones1999} behaves as a collisional fluid, whereas galaxies are effectively collisionless. 
Dark matter is assumed to be collisionless in the standard cold dark matter (CDM) paradigm, but this assumption is testable. 
For example, any self-interaction would decelerate the dark matter halos relative to the galaxies during a collision \citep[e.g.,][]{2015harvey, Bechtol2022arXiv, Jee2026}.
The distinct responses of these components make merging clusters powerful laboratories for probing the nature of dark matter \citep[e.g.,][]{Bechtol2022arXiv} and the physical mechanisms responsible for particle acceleration \citep[e.g.,][]{Brunetti2014, 2019vanweeren}.
No single observable, however, can separate these components since each traces only one of them, and their signatures are superposed along the line of sight.
Multiwavelength observations are therefore essential for disentangling these interactions, as they trace the different responses of galaxies, dark matter, the thermal ICM, and relativistic particles \citep[e.g.,][]{HyeongHan2024_a746, HyeongHan2025, HyeongHan2026arXiv, Finner2023, Finner2025, Rajpurohit2024, Rajpurohit2025, Stroe2025, Ahn2025}.

The contrasting dynamical responses of the cluster components can produce spatial offsets during a merger.
Hydrodynamic interactions decelerate and displace the gaseous ICM, whereas galaxies and dark matter continue along trajectories governed primarily by gravity.
The magnitude of the resulting ICM--dark matter separation depends on several factors, including the collision velocity, impact parameter, gas density, merger stage, and projection. 
Systems exhibiting extreme separations, such as the Bullet Cluster \citep{2006clowe,2008randall}, MACS~J0025.4$-$1222 \citep{2008bradac}, CIZA~J0107.7+5408 \citep{2016randall, Finner2023}, and Abell~56 \citep{2023wittman}, are commonly referred to as dissociative mergers and have played a key role in constraining the properties of dark matter \citep[e.g.,][]{2004markevitch,2015harvey,2025cha_bullet,Rihtarsic2026}.
Because these spatial offsets evolve throughout a collision, translating them into dark matter constraints requires independent knowledge of the merger geometry and evolutionary stage \citep[e.g.,][]{Wittman2018}.

Merger shocks provide a particularly direct probe of the merger dynamics \citep[e.g.,][]{Russell2010, Russell2012, White2015, Jee2015, Okabe2015, Golovich2017, Cho2022, Lee2022, Finner2025}.
Axial or bow shocks propagate ahead of the advancing cores \citep[e.g.,][]{2002markevitch, Zhang2019}, whereas equatorial shocks expand approximately perpendicular to the merger axis \citep{2018ha, Gu2019NatAs, Chen2025}. 
Radio relics provide a direct probe of merger shocks. 
Double-radio-relic systems are particularly useful because the relics approximately trace shocks propagating in opposite directions from the merger center and frequently indicate a merger axis close to the plane of the sky \citep[e.g.,][]{vanWeeren2011b}.
These diffuse synchrotron sources are thought to be powered by relativistic electrons accelerated or re-accelerated at shock fronts \citep[for a review, see][]{2019vanweeren}.
Under the diffusive shock acceleration framework \citep{Kardashev1962, Blandford1987}, the shock Mach number can be inferred from the injection spectral index or, under assumptions of a stationary planar shock, from the integrated spectral index \citep[e.g.,][]{Botteon2020}.
Combined with an estimate of the upstream sound speed, the Mach number can then be used to infer the shock propagation velocity, thereby providing a chronometer of cluster mergers \citep[e.g.,][]{Jee2026}.

In X-ray observations, shock compression produces an edge in the surface-brightness (SB) distribution, while the associated temperature jump can be measured spectroscopically. 
When interpreted using the Rankine--Hugoniot jump conditions \citep{Landau1959}, these measurements constrain the shock Mach number and propagation velocity. 
The X-ray- and radio-derived Mach numbers, however, can differ because the two methods weight the shock structure differently and are affected differently by projection \citep[e.g.,][]{Botteon2020, Wittor2021, Whittingham2026}.
Comparing the shock locations and Mach numbers inferred from X-ray and radio observations therefore provides constraints on the three-dimensional shock structure, merger geometry, and time since pericenter passage \citep[e.g.,][]{2024lee, Lee2026}.
Such comparisons require reliable X-ray measurements, which are challenging in cluster outskirts where the emissivity scales proportional to the square of the gas density and therefore demand deep observations with sufficient signal to noise.

MACS~J1752.0+4440 ($z=0.365$; hereafter MACSJ1752) and ZwCl~1856.8+6616 ($z=0.304$; hereafter ZWCL1856) are double-radio-relic merging clusters.
The two systems provide an informative comparison because they share several properties. 
Each is an equal-mass merger with two dominant weak-lensing (WL) mass peaks located near the corresponding brightest cluster galaxies (BCGs) and X-ray SB peaks \citep{Finner2021}. 
Their merger axes are inferred to lie close to the plane of the sky, as suggested by their double-relic geometries, radio polarization \citep{Ensslin1998, vanWeeren2012, 2012bonafede, Jones2021, Chiesa2026arXiv}, and small line-of-sight velocity differences between the spectroscopically identified subclusters \citep{vanWeeren2012, 2012bonafede, 2014degasperin, 2019agolovich}. 
Although their projected relic separations\footnote{The relic separation is defined as the distance between the centers of the two radio-relic shock fronts \citep{Finner2021}.} are similar, approximately 2.3~Mpc in MACSJ1752 and 2.1~Mpc in ZWCL1856, their total masses differ by nearly an order of magnitude.
MACSJ1752 has $M_{200c}=14.7^{+3.8}_{-3.3}\times10^{14}~M_{\odot}$\footnote{$M_{\Delta}$ corresponds to the mass enclosed within a sphere whose mean density equals $\Delta$ times the critical density of the background universe, $\rho_c(z)$, at the cluster redshift, where $\Delta$ is the density contrast.}, whereas ZWCL1856 has $M_{200c}=2.4^{+0.9}_{-0.7}\times10^{14}~M_{\odot}$ \citep{Finner2021}.
Thus, the similar physical relic separations do not necessarily indicate the same stage of shock or merger evolution, motivating an independent assessment of the merger stage through the X-ray analysis.

MACSJ1752 was discovered in the MAssive Cluster Survey \citep{Ebeling2001}. 
Its northeastern (NE) and southwestern (SW) radio relics have integrated spectral indices of $\alpha_{\rm NE}=-1.16\pm0.03$ and $\alpha_{\rm SW}=-1.10\pm0.05$, respectively \citep{vanWeeren2012}. 
The spectral indices steepen toward the cluster center, and the polarization fraction reaches $\sim$40\% at the relic fronts \citep{2012bonafede}. 
Recent radio measurements yield Mach numbers of $\mathcal{M}_{\rm NE,radio}=3.1\pm0.1$ and $\mathcal{M}_{\rm SW,radio}=3.2\pm0.1$ \citep{Chiesa2026arXiv} for the NE and SW relics, respectively.
However, previous X-ray observations \citep{Finner2021} did not reveal clear SB or temperature discontinuities associated with either relic, largely because of limited photon statistics.

ZWCL1856, also known as PSZ2~G096.88+24.18, was detected through the Sunyaev--Zel'dovich effect \citep{Sunyaev1972} by \textit{Planck} \citep{Planck2011,Planck2016}. 
Two Mpc-scale radio relics were identified on opposite sides of the cluster by the Westerbork Synthesis Radio Telescope observations at north and south of the cluster \citep{2014degasperin}. 
LOFAR and VLA observations yield radio Mach numbers of $\mathcal{M}_{\rm N,radio}=2.5\pm0.2$ and $\mathcal{M}_{\rm S,radio}=2.3\pm0.2$ \citep{Jones2021}, although no corresponding SB discontinuities were detected in the older, shallow \textit{Chandra} observations.
Recently, \citet{Tumer2026} reported temperature jumps from deep ($\sim$300~ks) \textit{NuSTAR} observations corresponding to $\mathcal{M}_{\rm N,X}=3.90^{+1.64}_{-0.85}$ and $\mathcal{M}_{\rm S,X}=2.36^{+0.58}_{-0.46}$.

Although Paper I \citep{Finner2021} established the weak-lensing mass distributions and their relation to the galaxies, radio relics, and X-ray emission, the relatively shallow X-ray observations available at the time prevented a detailed characterization of the gas morphology and thermodynamic structure.
The substantially deeper X-ray observations analyzed here make it possible to investigate these features in greater detail and to connect the observed ICM structures to the merger dynamics \citep[e.g.,][]{Markevitch2006, Simionescu2012, Lyskova2019, Churazov2021, Zhuravleva2023, Rajpurohit2024, Stroe2025}.
This enables us to address a central question: how can mergers with similar radio-relic and projected mass configurations develop such different gas morphologies, and what do these differences reveal about their impact parameters, assembly histories, and times since pericenter passage?

In this work, we present a comparative X-ray analysis of MACSJ1752 and ZWCL1856 to characterize their thermodynamic structures, search for X-ray signatures of merger shocks, and reconstruct their merger histories. 
We use deep XMM-\textit{Newton} observations of MACSJ1752 and \textit{Chandra} observations of ZWCL1856, increasing the available exposure times from $\sim$10 to $\sim$135~ks and from $\sim$37 to $\sim$209~ks, respectively. 
Section~\ref{sec:observations} describes the observations and data-reduction procedures, while Section~\ref{sec:analysis} presents the SB-profile modeling and instrument-specific spectral-analysis methods. 
In Section~\ref{sec:results}, we present searches for X-ray discontinuities associated with the radio relics, spatially resolved temperature and thermodynamic maps, and analyses of the extended X-ray structures in both clusters. 
In Section~\ref{sec:discussion}, we discuss our results in combination with the available radio, weak-lensing, and galaxy data and compare the observed systems with analogs identified in cosmological simulations. 
We use these comparisons to refine their merger scenarios and assess the suitability of double-radio-relic clusters for dark matter studies. 
We summarize our findings in Section~\ref{sec:conclusion}.

We adopt $H_0=70~\rm km~s^{-1}~Mpc^{-1}$, $\Omega_{M}=0.3$, $\Omega_{\Lambda}=0.7$ in a flat $\Lambda$CDM cosmology throughout this paper.
At the redshift of MACSJ1752 ($z=0.365$), the plate scale is $5.1~\rm kpc~arcsec^{-1}$.
At the redshift of ZWCL1856 ($z=0.304$), the plate scale is $4.5~\rm kpc~arcsec^{-1}$.
Reported uncertainties correspond to the 68 per cent confidence level, unless stated otherwise.

\begin{figure*}
    \centering
    \includegraphics[width=1\linewidth]{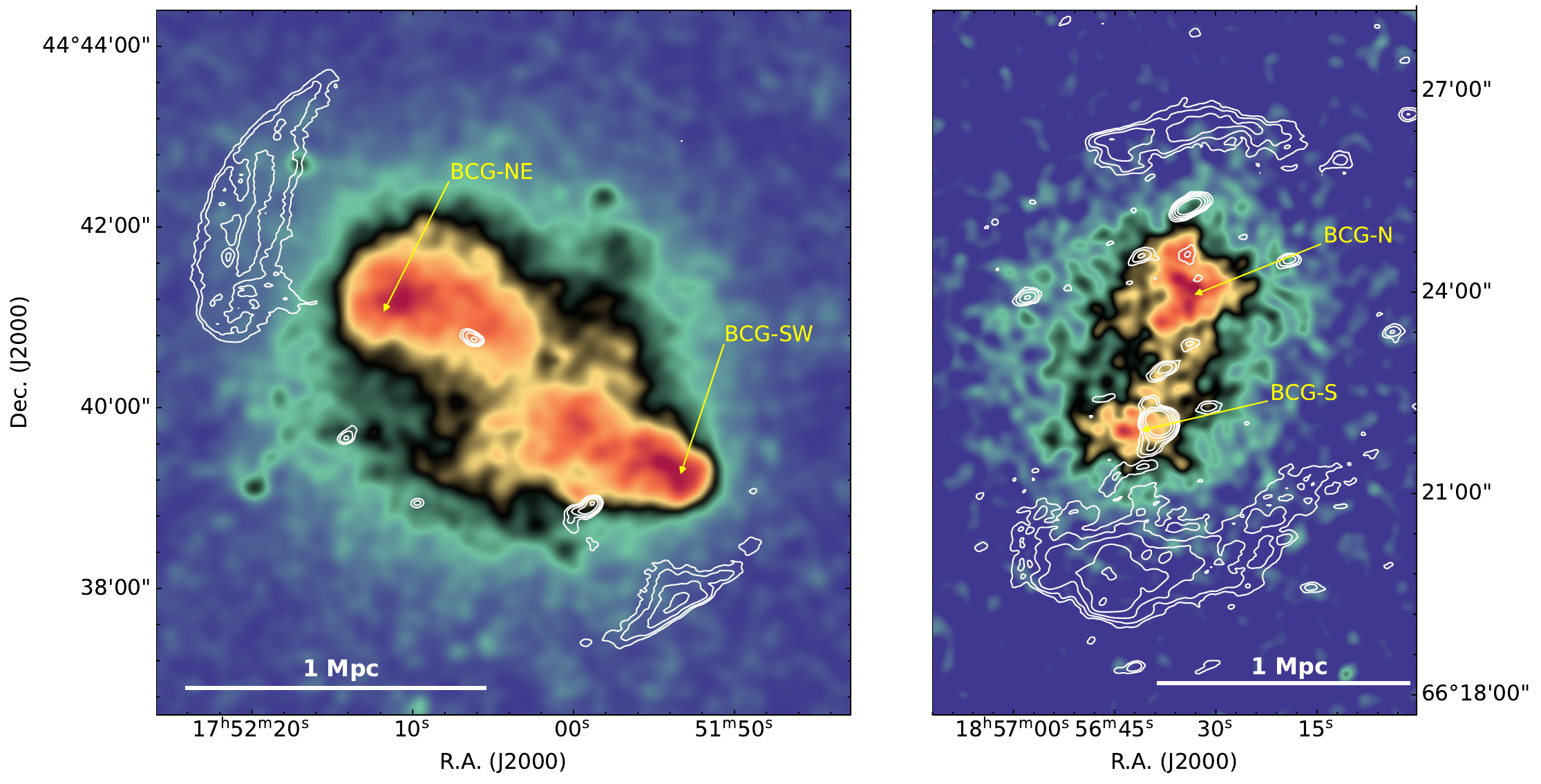}
    \caption{Background-subtracted, point-source-subtracted, and exposure-corrected 0.5--2.0~keV X-ray SB maps of MACSJ1752 (left; XMM-{\it Newton}) and ZWCL1856 (right; {\it Chandra}). 
             Both maps were smoothed with a Gaussian kernel of $\sigma=4\farcs5$. 
             White contours show the diffuse radio emission observed with LOFAR at 144~MHz for MACSJ1752 \citep{Chiesa2026arXiv} and at 140~MHz for ZWCL1856 \citep{Jones2021}. 
             In both panels, the contours are drawn at $[3,6,12,24,48]\sigma$, where the rms noise levels are $\sigma_{\rm MACSJ1752}=280~\mu{\rm Jy~beam^{-1}}$ and $\sigma_{\rm ZWCL1856}=127~\mu{\rm Jy~beam^{-1}}$, respectively. 
             The radio contours trace the double radio relics in each system. 
             Yellow arrows indicate the BCGs associated with the subclusters. 
             MACSJ1752 exhibits prominent extended X-ray tails between and behind the subclusters, whereas ZWCL1856 shows laterally displaced X-ray-emitting gas extending beyond the BCGs.
            }
    \label{fig:main}
\end{figure*}

\section{Observations and Data Reduction} \label{sec:observations}

\subsection{XMM-Newton observations}

MACSJ1752 was observed three times by XMM-{\it Newton} (ObsID: 0650383401 -- PI: S. Allen; ObsID: 0903130101 and 0903130201 -- PI: K. HyeongHan) with a total exposure time of $\sim$148~ks.
We processed the observation data files using the XMM-{\it Newton} Extended Source Analysis Software (XMM-ESAS; SAS version 18.0.0; \cite{2004gabriel}) following the standard procedures described in the XMM-ESAS Cookbook\footnote{http://heasarc.gsfc.nasa.gov/docs/xmm/esas/cookbook/}.
The calibrated event files were generated by running the {\tt emchain} and {\tt epchain} tasks.
The events with PATTERN$>$12 for the MOS data and with PATTERN$>$4 for the PN data were excluded.
We applied the expression FLAG==0 to remove bright pixels and hot columns and performed PN out-of-time correction.
High background signals due to solar flares were filtered by the {\tt mos-filter} and {\tt pn-filter} tools.
We did not include the anomalous-state CCDs \citep{Kuntz2008} during the analysis.
The final MOS1, MOS2, and PN clean exposure times are 135.8, 135.3, and 108.1 ks, respectively (Table~\ref{tab:xray_obs}).

Point sources detected with the \texttt{edetect\_chain} task were masked. 
For visualization in Figure~\ref{fig:main}, the masked regions were filled using the CIAO task \texttt{dmfilth}. 
However, \texttt{dmfilth} did not adequately reconstruct the diffuse emission beneath the source at $\mathrm{R.A.}=17{:}52{:}01.08$ and $\mathrm{Dec.}=+44{:}40{:}31.53$, because it lies within the spatially varying cluster emission. 
We therefore modeled this source with a two-dimensional $\beta$ profile with an exponential cutoff and subtracted the best-fit model from the displayed image. 
These modifications were performed solely for cosmetic purposes, and the resulting image was not used in any quantitative analysis.

The instrumental background was modeled separately for each EPIC detector using the XMM-ESAS \texttt{mos\_back} and \texttt{pn\_back} tasks, adopting the same energy bands and detector selections as for the science data.
The resulting background products were transformed into sky coordinates using \texttt{rot-im-det-sky} with the same spatial masks as the science images.
Residual soft-proton contamination was modeled using the \texttt{proton\_scale} and \texttt{proton} tasks, with its normalization constrained using an off-source annular region.
The quiescent-particle and residual soft-proton background components were then combined with the EPIC imaging products using the XMM-ESAS \texttt{comb} task.
To constrain the astrophysical sky background, we jointly fitted the EPIC spectra extracted from an off-source annulus at $R>3$~Mpc and a ROSAT All-Sky Survey (RASS) spectrum sampling the surrounding field. 
The best-fit sky model was then used to estimate the residual 0.5--2 keV sky-background level for the SB analysis in Section~\ref{subsec:sb_profile}.
The left panel of Figure~\ref{fig:main} shows the smoothed ($\sigma=4\farcs5$) soft band (0.5--2 keV) image with a binning of 40 physical pixels (corresponding to $2''$).

\subsection{Chandra observations}

ZWCL1856 was observed for a total of 259~ks with Chandra X-ray Observatory (ObsID: 19752 -- PI: R. Kraft; ObsID: 26933, 27204, 27205, 27206, 27207, 27208, 27209, 29201, 29293, 29294 -- PI: K. Finner) in VFAINT Mode. 
The cluster was positioned on the ACIS-I CCD with rotations applied between observations to eliminate the impact of chip gaps. 
We applied a modified version of the \textit{Chandra} data reduction/analysis methodology described in \citet{silich2024.apj} using \texttt{CIAO} version 4.16 \citep{ciao}. 
We calibrated the raw ACIS-I data for all ZWCL1856 ObsIDs using the CalDB version 4.11.3 with \texttt{chandra\_repro}. 
For each dataset, we performed an exposure correction (\texttt{fluximage}), point-source identification (\texttt{wavdetect}) and exclusion, and light-curve filtering (\texttt{deflare}). 
The total clean exposure time is $\simeq209$ ks. 
From the clean event files, we generated soft (0.5--2~keV) and broad (0.5--7~keV) band flux and counts maps (\texttt{flux\_obs}). 
Blank-sky background event files were generated for each observation using the CIAO \texttt{blanksky} task and reprojected to match the corresponding science event file. 
The blank-sky background was normalized independently for each ACIS CCD using the ratio of counts in the particle-dominated 9–12 keV band. 
Background images in the 0.5–2.0 keV band were then generated with \texttt{blanksky\_image} using the same spatial grid as the science images.
For visualization, we masked the regions around detected point sources and filled the holes using the CIAO task \texttt{dmfilth}.
The smoothed ($\sigma=4\farcs5$) soft band (0.5--2 keV) X-ray SB map of ZWCL1856 with a binning of 8 physical pixels (corresponding to $4''$) is shown in the right panel of Figure \ref{fig:main}.

\section{Analysis \& Methods} \label{sec:analysis}

Although the two clusters were observed with different telescopes, XMM-{\it Newton} and Chandra, we describe the common analysis steps here and note instrument-specific difference where applicable.
When the method differs, we explicitly state the telescope to which it is being applied.

\subsection{Surface brightness profile} \label{subsec:sb_profile}

For both clusters, we extracted and modeled the X-ray SB profiles across the radio relics using the {\tt pyproffit} package \citep{Eckert2020}. 
The profiles were extracted from the point-source-masked, background-subtracted 0.5--2.0~keV images using sectors guided by the morphology and location of the radio relics (Figures~\ref{fig:macsj1752_profile} and \ref{fig:zwcl_sb_profile}) with the corresponding region definitions listed in Table~\ref{tab:sb_regions}.
We adopted radial bin widths of $7\arcsec$ for XMM-{\it Newton} and $5\arcsec$ for Chandra to account for the point-spread functions (PSFs), and the profiles were regrouped to have a minimum signal-to-noise ratio of 5 in each radial bin. 
The SB profiles for the NE and SW regions of MACSJ1752 are shown in the right panels of Figure \ref{fig:macsj1752_profile} and those of ZWCL1856 are shown in the right panels of Figure \ref{fig:zwcl_sb_profile}.

\begin{table*}
\centering
\caption{Regions used for the SB profile analysis.}
\label{tab:sb_regions}

\setlength{\tabcolsep}{6pt}
\begin{tabular}{cccccccc}
    \hline\hline
    \multirow{2}{*}{Cluster} &
    \multirow{2}{*}{Region} &
    R.A. & Dec. & $\theta_1$ & $\theta_2$
    & $R_{\rm in}$ & $R_{\rm out}$ \\
    & & (deg) & (deg) & (deg) & (deg)
      & (arcmin) & (arcmin) \\
    \hline
    \multirow{2}{*}{MACSJ1752}
        & NE & 268.039331 & 44.692226 & 87 & 202 & 1.1 & 5 \\
        & SW & 268.014125 & 44.674730 & 277 & 331 & 2.2 & 4 \\
    \hline
    \multirow{2}{*}{ZWCL1856}
        & N  & 284.127054 & 66.380875 & 66 & 119 & 2 & 5 \\
        & S  & 284.177287 & 66.401495 & 243 & 324 & 2.5 & 5.8 \\
    \hline
\end{tabular}
\tablecomments{R.A. and Dec. give the centers of the circular sectors used to extract the SB profiles. The angles $\theta_1$ and $\theta_2$ define the angular boundaries of each sector, while $R_{\rm in}$ and $R_{\rm out}$ specify its inner and outer radii, respectively.}
\end{table*}

To identify and characterize possible SB discontinuities, we fitted each profile with a projected broken power-law density model. 
We assume that emissivity variations are negligible, since the temperature dependence is weak at these temperatures and the abundance is not expected to change substantially across the shock front.
\begin{equation}
I(r) = I_0 \int F(\omega)^2 ~ d\ell + B,
\end{equation}
where $I_0$ is the normalization and $B$ is the sky background which was fixed to 0 as it is subtracted prior to the model fitting.
The three-dimensional density model is defined as
\begin{equation}
F(\omega) =
\begin{cases}
\omega^{-\alpha_1}, & \omega < r_f, \\
\dfrac{1}{C}\omega^{-\alpha_2}, & \omega \ge r_f ,
\end{cases}
\end{equation}
where $r$ is the projected radius from the adopted center, $\ell$ is the line-of-sight distance, and $\omega=\sqrt{r^2+\ell^2}$. 
The break radius $r_f$ represents the radial position of the discontinuity. 
The parameters $\alpha_1$ and $\alpha_2$ describe the power-law slopes in the downstream and upstream regions, respectively, and $C=n_{\rm e,post}/n_{\rm e,pre}$ is the electron density jump across the discontinuity. 
The XMM-{\it Newton} PSF was modeled as a King function, ${\rm PSF}(x)=\left[1+\left(x/r_0\right)^2\right]^{-\alpha}$, where $r_0=0\farcm0883981$ is the core radius and $\alpha=1.58918$ is the outer slope.
We chose not to apply the same PSF treatment to the \textit{Chandra} profiles because the {\it Chandra} PSF is substantially narrower than the adopted ($5\arcsec$) radial bins and the spatial scales of the fitted discontinuities.

\subsection{Spectral analysis} \label{subsec:spec}

X-ray spectroscopy provides information about the thermal state of the ICM.
Although an X-ray SB discontinuity traces a change in the projected gas density, its physical origin cannot be securely established without the corresponding temperature structure.
Spatially resolved spectral analyses  allow us to identify merger-induced shocks, cool cores, and thermally disturbed gas structures.
These motivate our spectral analyses of MACSJ1752 and ZWCL1856, for which the instrument-specific procedures are described below.

\subsubsection{MACSJ1752} \label{subsec:macsj1752_spec}
We performed an X-ray spectral analysis using {\tt XSPEC} ({\tt v12.11.1}). 
The EPIC spectra from two XMM-{\it Newton} observations, 0903130101 and 0903130201, were fitted jointly, while the observation 0650383401 was excluded from the analysis because of its high particle background. 
The joint spectral analysis included the MOS1, MOS2, and PN spectra from each observation, together with their corresponding detector-specific background spectra, redistribution matrix files (RMFs), and ancillary response files (ARFs). 
The source spectra were grouped to contain at least one count per channel and fitted using the C-statistic \citep{Cash1979}.
Throughout the analysis, we adopted the solar abundance table of \citet{Asplund2009} and the photoelectric absorption cross sections of \citet{Verner1996} when modeling Galactic absorption.
The XMM-{\it Newton} spectra were fitted over the 0.5--10.0~keV energy range. 
In addition, we included a RASS spectrum extracted from a surrounding annular region and fitted it over the 0.1--2.0~keV energy range to constrain the astrophysical X-ray background.

To account for both sky and instrumental contributions in the spectral fitting, we constructed a background model consisting of two separate components: the astrophysical X-ray background (AXB), which accounts for sky emission, and the non-X-ray background (NXB), which accounts for the instrumental particle background.
The AXB model is defined as
\begin{equation}
    {\tt apec}_{\rm LB}
    + {\tt phabs}\times
    \left(
    {\tt apec}_{\rm MW}
    + {\tt powerlaw}_{\rm CXB}
    \right),
\end{equation}
where the Local Bubble (LB) component was modeled as an unabsorbed thermal plasma with a fixed temperature of $kT=0.12$~keV. 
The Milky Way foreground emission was modeled as an absorbed thermal plasma with $kT=0.20$~keV, and the cosmic X-ray background (CXB) was modeled as an absorbed power law with a fixed photon index of $\Gamma=1.41$. 
The metal abundances and redshifts of the LB and Milky Way components were fixed to $1~Z_{\odot}$ and zero, respectively. 
The Galactic hydrogen column density was fixed to $N_{\rm H}=3.2\times10^{20}~{\rm cm^{-2}}$ \citep{HI4PI2016}.

The cluster emission was modeled with an absorbed thermal plasma component, ${\tt apec}_{\rm ICM}$, at MACSJ1752's redshift of $z=0.365$. 
The ICM abundance was fixed to $0.3~Z_{\odot}$, while the ICM temperature and normalization were allowed to vary. 
For the RASS spectrum, the ICM normalization was fixed to zero so that the RASS data constrained only the sky background components. 
The spectral shapes of the LB, Milky Way, and CXB components were tied between the XMM-{\it Newton} and RASS spectra.

The NXB was modeled explicitly for each detector and observation. 
For each EPIC spectrum, the NXB model consisted of fluorescent instrumental lines modeled with Gaussian components \citep{Su2017}, plus a broken power-law continuum.
The centroid energies of the Gaussian components were fixed to the expected detector line energies and their line widths were constrained to be smaller than 0.3~keV. 
The particle continuum was modeled with a broken power law whose break energy was fixed at 3~keV. 
These NXB components were folded through the detector RMFs but not through the ARFs, because they represent instrumental background. 
The NXB parameters were fixed during the final source fit.

\subsubsection{ZWCL1856}

We analyzed the \textit{Chandra} spectra using \texttt{Sherpa}. 
For each spatial region and ObsID, we extracted source spectra from the cleaned event files and corresponding background spectra from the blank-sky event files. 
The blank-sky data include the time-averaged particle-induced instrumental background and an average unresolved astrophysical X-ray background and foreground, and were normalized to each observation using the particle-dominated 9--12~keV count rate.
All identified point sources were excluded from the spectral extractions. 
The spectra were grouped to contain at least 25 counts per bin, allowing the use of $\chi^2$ statistics.
For each spatial region, the blank-sky-background-subtracted spectra from all ObsIDs were fitted jointly over the 0.5--7.0~keV energy range.

The source emission was modeled as a single-temperature collisional-ionization-equilibrium plasma using \texttt{apec} \citep{Smith2001,Foster2012}, modified by Galactic absorption using \texttt{tbabs} \citep{tbabs}. 
The plasma temperature and normalization were allowed to vary, while the metallicity was fixed to $Z=0.3~Z_{\odot}$ using the abundance table of \citet{angrabunds}. 
The Galactic hydrogen column density and cluster redshift were fixed to $N_{\rm H}=4.8\times10^{20}~{\rm cm^{-2}}$ \citep{HI4PI2016} and $z=0.304$, respectively.

\section{Results} \label{sec:results}

\subsection{Surface brightness discontinuities} \label{subsec:macsj1752_shock}

\begin{figure*}
    \centering
    \includegraphics[width=1\linewidth]{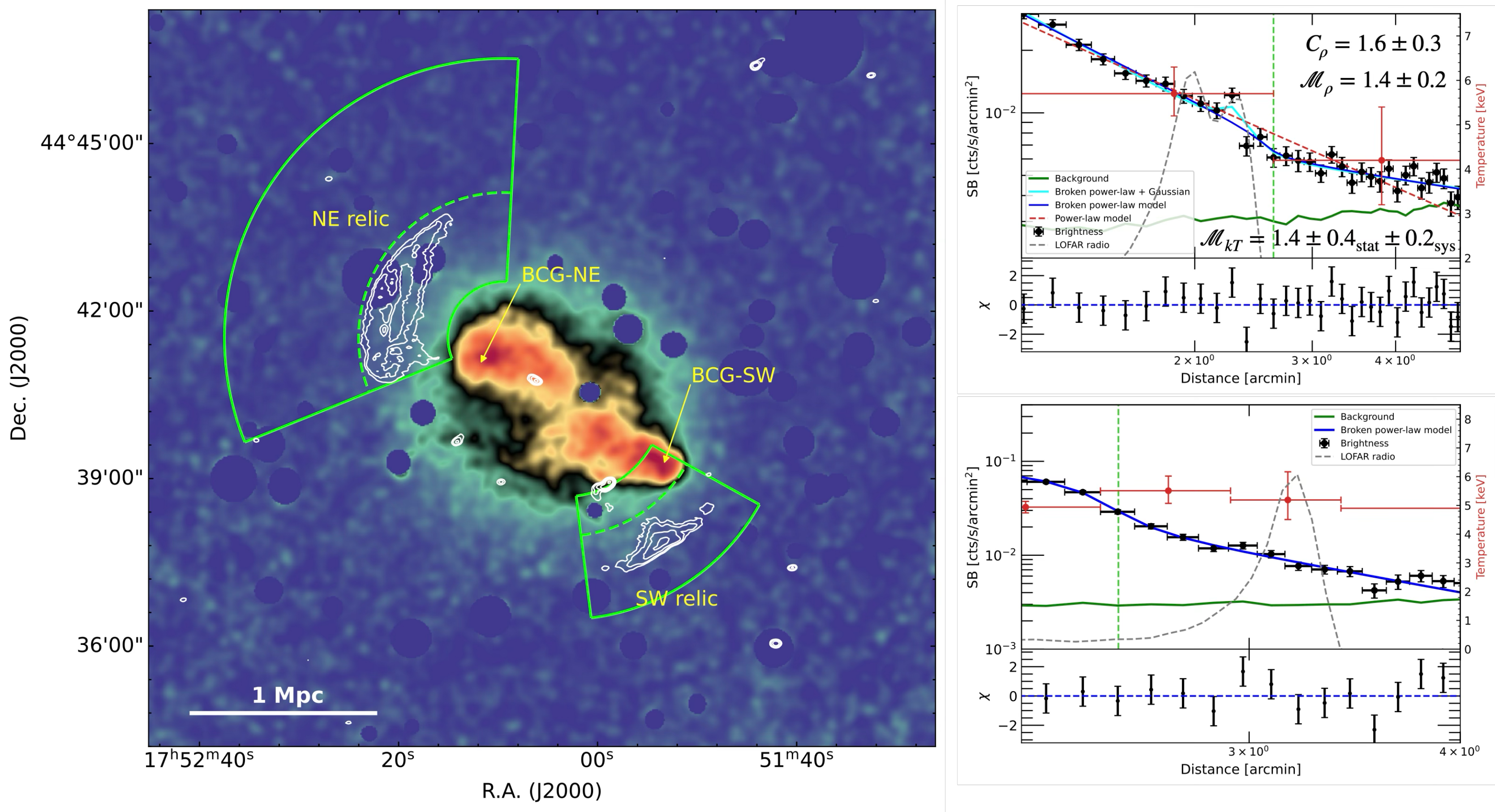}
    \caption{X-ray SB profiles in the 0.5--2~keV band along the putative shock directions guided by the radio relics.
             Left: Same as in Figure~\ref{fig:main} for MACSJ1752 where the point sources are masked.
             Green dashed lines indicate the best-fit break radii.
             Top right: The X-ray background-subtracted SB profile is extracted in the NE radio relic region.
             The best-fit broken power-law model ($\chi^2_{red}=1.1$; d.o.f. = 29) is shown in blue while the best-fit single power-law model ($\chi^2_{red}=2.7$; d.o.f. = 31) is shown as a red dashed line.
             The cyan line is the best-fit broken power-law + Gaussian model ($\chi^2_{red}=1.1$; d.o.f. = 26) where the residual is shown below (see Appendix~\ref{app:IC} for more details).
             The temperature profile is shown in red with $1\sigma$ uncertainties.
             The horizontal error bars indicate the bin widths.
             The gray dotted curve represents the LOFAR radio SB profile in arbitrary unit within the same region that highlights the radio relic.
             The green solid line indicates the background signal.
             The Mach numbers derived from the density ($\mathcal{M}_{\rho}$) and temperature ($\mathcal{M}_{kT}$) jumps are indicated.
             Bottom right: Same as the upper panel, but the X-ray SB profile is extracted in the SW relic region. 
             The prominent SB discontinuity is caused by the cold front of the SW subcluster.
            }
    \label{fig:macsj1752_profile}
\end{figure*}

Radio relics are commonly interpreted as synchrotron emission from relativistic electrons accelerated by merger shocks \citep[e.g.,][]{2019vanweeren}. 
Therefore, identifying jumps in gas density and temperature in X-ray observations that are spatially coincident with the radio relics provides independent confirmation of a shock origin and an estimate of the shock strength \citep{2007markevitch}. 
It is expected that a shock should present a temperature increase from the pre- to post-shock region. 
As merger shocks propagate away from the cluster center, the post-shock regions lie on the cluster-facing sides of relics, whereas the pre-shock regions lie immediately beyond their outer edges, corresponding to the expected downstream and upstream regions, respectively. 
We thus examine the X-ray SB profiles and temperature measurements across the relic directions and compare the inferred X-ray Mach numbers with those suggested by the radio data.

\textbf{MACSJ1752} hosts radio relics in the NE and SW directions with Mach numbers of $\mathcal{M}_{\rm NE}=3.1\pm0.1$ and $\mathcal{M}_{\rm SW}=3.2\pm0.1$ \citep{Chiesa2026arXiv}, respectively. 
Figure~\ref{fig:macsj1752_profile} shows the X-ray SB profile along the NE and SW radio relics of MACSJ1752 and both the power-law and broken power-law fits.

For the NE relic, the best-fit broken power-law model ($\chi^2_{red}=1.1$; degrees of freedom = 29) is preferred over the power-law fit ($\chi^2_{red}=2.7$; d.o.f. = 31). 
The inner and outer power-law slopes are $1.32\pm0.09$ and $0.72\pm0.08$, respectively. 
The best-fit break radius ($2\farcm6$) and the front edge of the NE relic are in remarkable spatial agreement, suggesting the presence of a merger-shock-induced X-ray SB discontinuity.

The resulting compression factor is $C_{\rho} = 1.6 \pm 0.3$ and the corresponding Mach number, assuming Rankine--Hugoniot jump conditions \citep{Landau1959}, is $\mathcal{M}_{\rho}=1.4\pm0.2$. 
In addition, we test for an excess potentially associated with inverse Compton (IC) emission by adding a Gaussian component to the broken-power-law model (cyan). 
We discuss this analysis further in Appendix~\ref{app:IC}

We examined the ICM temperature from upstream to downstream by extracting spectra and fitting them as described in Section~\ref{subsec:macsj1752_spec}.
The temperature measurements of the pre- and post-shock regions of the NE relic are $kT=4.2^{+1.2}_{-1.0}$ and $kT=5.7^{+0.6}_{-0.5}$, respectively.
Assuming the Rankine--Hugoniot temperature-jump condition for an ideal gas with $\gamma=5/3$, we obtain a Mach number of $\mathcal{M}_{kT}=1.4\pm0.4_{\rm stat}\pm0.2_{\rm sys}$ (see Appendix~\ref{app:sys} for the systematic uncertainty).

The density discontinuity implies $\mathcal{M}_{\rho}=1.4\pm0.2$, whereas the temperature jump provides only a marginal constraint, with $\mathcal{M}_{kT}=1.4\pm0.4_{\rm stat}\pm0.2_{\rm sys}$.
Conservatively, the temperature measurement therefore provides an upper limit of approximately $\mathcal{M}_{kT}\lesssim2.3$ at the $2\sigma$ level.
Even this upper limit remains well below the radio-derived $\mathcal{M}_{\rm NE}=3.1\pm0.1$ \citep{Chiesa2026arXiv}.
This radio--X-ray Mach-number discrepancy is commonly found in radio relic shocks \citep{Botteon2020}. 
One possible explanation is that the radio and X-ray measurements probe different effective averages of a non-uniform shock surface \citep{Wittor2021, Whittingham2026}. 
The radio emission is dominated by the highest-Mach-number regions, which produce most of the synchrotron power, whereas the X-ray SB and temperature jumps are averaged over the broader thermal ICM structure and are more strongly affected by projection. 
Although the merger axis of MACSJ1752 is inferred to lie close to the plane of the sky \citep{2012bonafede, 2017golovich}, residual projection effects may arise from the curvature and inhomogeneity of the shock front \citep[e.g.,][]{2024lee, Lee2025}. 
Mixing of pre- and post-shock gas along the line of sight can reduce the observed density and temperature contrasts and thereby bias the X-ray-derived Mach number low \citep{Hong2015, Zhang2019}.
Thus, the lower X-ray values, $\mathcal{M}_{\rho}=1.4\pm0.2$ and $\mathcal{M}_{kT}=1.4\pm0.4_{\rm stat}\pm0.2_{\rm sys}$, are consistent with the idea that the radio relic preferentially highlights the stronger portions of an intrinsically inhomogeneous merger shock.

At the SW relic front, we follow the same procedure as for the NE relic but find no coherent discontinuity in either the X-ray SB or temperature extending across the full width of the relic. 
The best-fit broken-power-law model shown in Figure~\ref{fig:macsj1752_profile} instead traces the cold front ($\chi^2_{red}=1.5$; d.o.f. = 10; $C_{\rho}=2.5\pm0.2$) associated with the core of the SW subcluster. 
When the extraction region is tailored to follow the morphology of this edge (R.A. = 268.014125, Dec. = 44.660980, $R_{in}=1\farcm8$, $R_{out}=3\arcmin6$, $\theta_1=336^\circ$ and $\theta_2=356^\circ$), the inferred best-fit ($\chi^2_{red}=0.9$; d.o.f. = 11) density compression becomes $C_{\rho}=5.2\pm0.8$.
This large inferred compression, $(C_{\rho}\sim5$, motivates an independent consistency check using the temperature measurement across the edge. 
For a pressure-continuous cold front \citep[e.g.,][]{Mazzotta2002, 2007markevitch}, the temperature contrast should approximately compensate for the density jump. 
Using regions divided at the cold front ($r_f=2\farcm3$), we find $kT_{in}=4.0\pm0.3$~keV and $kT_{out}=4.9^{+1.2}_{-0.9}$~keV, giving a temperature ratio of only $\sim$1.2. 
The remaining discrepancy may arise from projection and line-of-sight mixing, which can reduce the apparent temperature contrast while biasing the density jump inferred under the assumed geometry \citep[e.g.][]{2007markevitch, Sanders2014, Werner2016}.

\begin{figure*}
    \centering
    \includegraphics[width=1\linewidth]{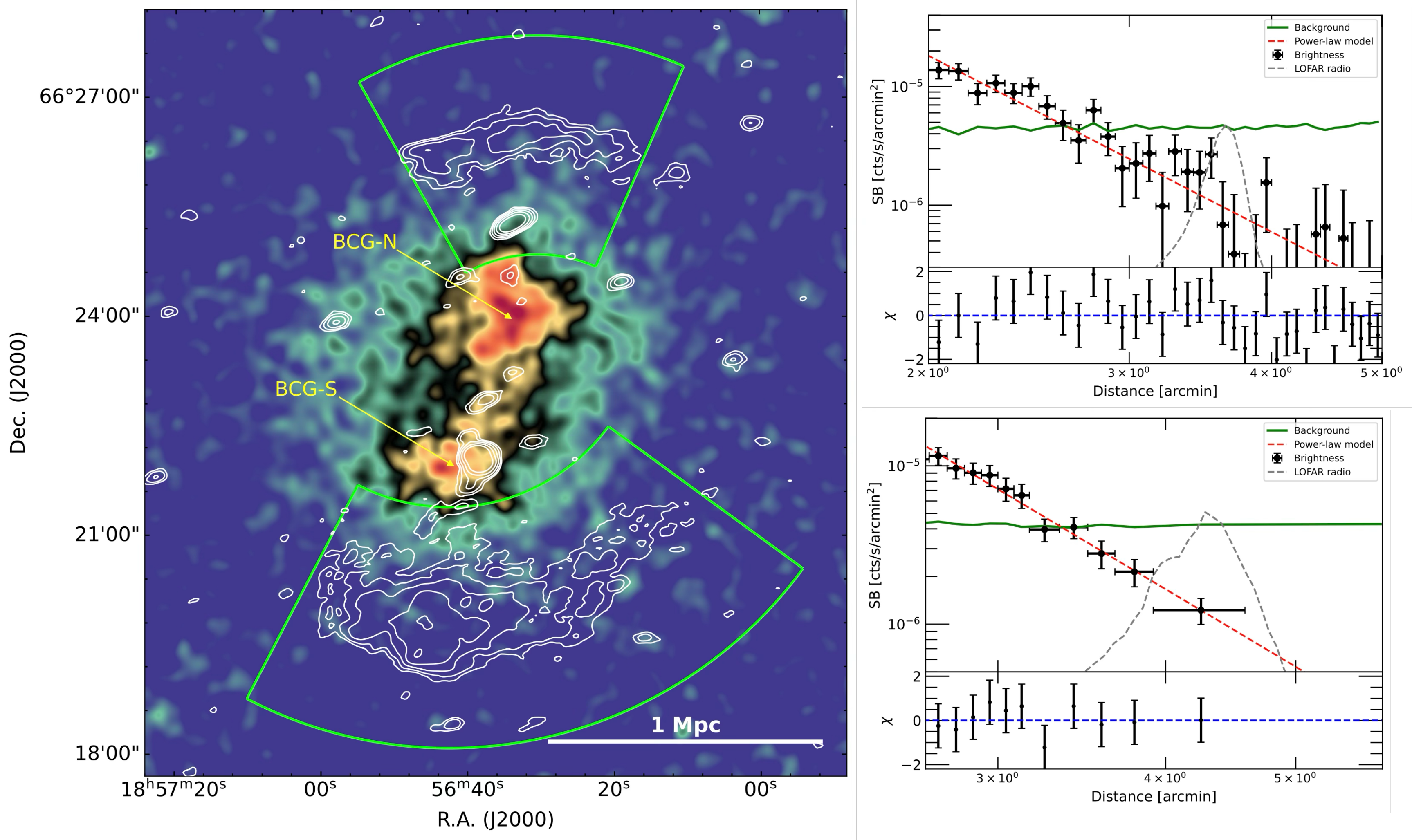}
    \caption{Same as Figure~\ref{fig:macsj1752_profile}, but for ZWCL1856. 
             The top and bottom panels show the X-ray background-subtracted SB profiles across the northern and southern radio relics, respectively. 
             The power-law fits yield $\chi^2_{\rm red,N}=1.3$ (d.o.f. = 32) and $\chi^2_{\rm red,S}=0.5$ (d.o.f. = 7). 
             Because the available counts are insufficient to achieve a minimum signal-to-noise ratio of 5 while retaining useful spatial resolution, the northern SB profile is shown in fixed $5\arcsec$ bins.
             We do not find an X-ray SB discontinuity at the location of either radio relic.
            }
    \label{fig:zwcl_sb_profile}
\end{figure*}

We examined an SB profile extracted from a narrower sector designed to exclude the cool core ($\theta_1=277^\circ$ and $\theta_2=306^\circ$) because emission from the southern cool core may contaminate the downstream region. 
Although the broken-power-law model provides an acceptable fit ($\chi^2_{red}=1.2$; d.o.f. = 25) and places the break near the SW relic, the amplitude of the density jump is not constrained.
We also note a relatively flat SB distribution at larger radii beyond the SW relic, which is examined in more detail in Appendix~\ref{app:excess}.

\textbf{ZWCL1856} hosts radio relics in the N and S with Mach numbers $\mathcal{M}_{\mathrm{N,radio}}=2.5\pm0.2$ and $\mathcal{M}_{\mathrm{S,radio}}=2.3\pm0.2$, respectively \citep{Jones2021}. 
Figure~\ref{fig:main} presents the X-ray SB map (0.5--2~keV) of ZWCL1856 with LOFAR 140~MHz radio contours \citep{Jones2021}. 
The southern subcluster has a distinct X-ray peak that lies east of the BCG. 
The southern edge of the X-ray emission is arc-shaped with an abrupt drop in brightness that corresponds to the cold front identified in \citet{Finner2021} (also see Figure~\ref{fig:zwcl_sb_profile}).
From the southern peak, the X-ray emission trails off toward the cluster center. 
The northern subcluster has an X-ray brightness peak that is offset by $\sim$60~kpc to the north of the BCG and a broader distribution of bright emission (Figure~\ref{fig:zwcl_tail}). 
The northern edge of the X-ray emission is triangular, but it is difficult to define clearly given the low counts.

To assess possible density jumps associated with the radio relics, we extract X-ray SB profiles across sectors defined by the relic geometry. 
To minimize contamination from the cold front, the profiles are measured only beyond its location. 
We find no clear SB discontinuity at either radio relic, and single power-law fits provide statistically adequate descriptions of the profiles, with $\chi^2_{\rm red,N}=1.3$ (d.o.f. = 32) and $\chi^2_{\rm red,S}=0.5$ (d.o.f. = 7). 
The photon statistics are also insufficient to obtain reliable temperature measurements from separate upstream and downstream regions at the relic fronts.
The limited sensitivity of the {\it Chandra} observations, together with the relatively low mass of the cluster, may hinder the detection of weak shocks.
We therefore refrain from concluding that no X-ray shocks are present. 
A more detailed characterization of the SB and temperature discontinuities, incorporating joint \textit{NuSTAR} and \textit{Chandra} spectral fitting across multiple energy bands, will be presented in a forthcoming paper.

\subsection{Projected thermodynamic maps} \label{subsec:thermo}

\subsubsection{MACSJ1752}

\begin{figure*}
    \centering
    \includegraphics[width=1\linewidth]{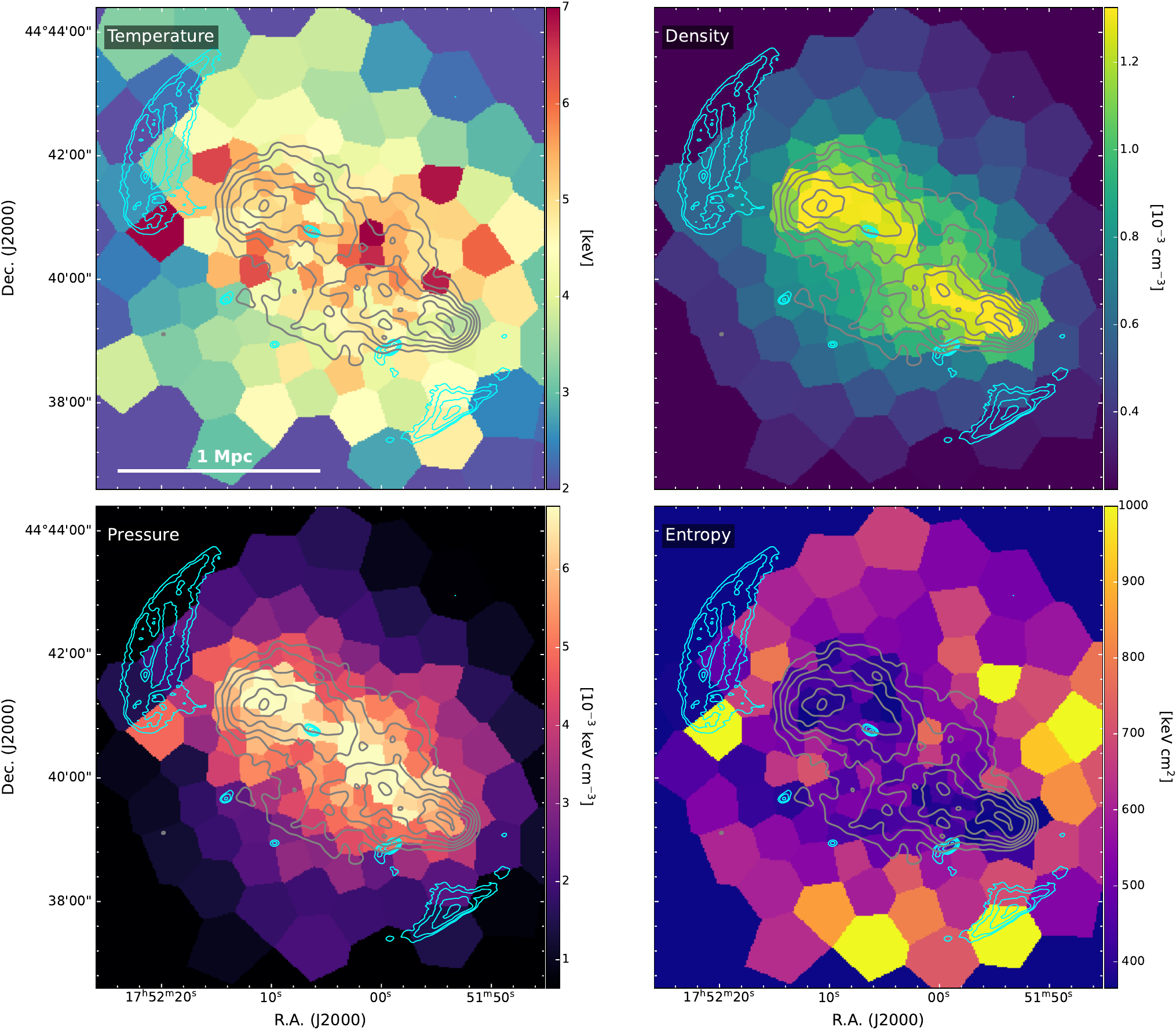}
    \caption{Projected thermodynamic maps of MACSJ1752. 
             From upper left to lower right, the panels show the temperature, electron density, pseudo-pressure, and pseudo-entropy maps. 
             Gray contours trace the soft-band X-ray SB at seven linearly spaced levels between $1.0\times10^{-5}$ and $3.0\times10^{-5}$~cts~s$^{-1}$. 
             Cyan contours show the LOFAR 144~MHz emission at $[3,6,12,24,48]\sigma$, where $\sigma=280~\mu{\rm Jy~beam^{-1}}$. 
             }
    \label{fig:macsj1752_thermo}
\end{figure*}

We constructed the X-ray temperature map by applying the same spectral-fitting procedure (Section~\ref{subsec:macsj1752_spec}) to spatially resolved spectra. 
We tessellated a point-source-subtracted, soft-band X-ray SB map using weighted Voronoi binning \citep{Cappellari2003, Diehl2006}. 
Each bin was required to have a signal-to-noise ratio of 30, which yields a temperature uncertainty $\lesssim20\%$ for most regions. 
For each Voronoi bin, we extracted the corresponding MOS1, MOS2, and PN spectra from the two XMM-{\it Newton} observations and fitted them jointly with the background and source models described in Section~{\ref{subsec:macsj1752_spec}}.
During these fits, the ICM temperature and normalization were left free, while the abundance and redshift were fixed to $0.3~Z_{\odot}$ and $z=0.365$, respectively. 
The resulting best-fit temperatures were assigned to the individual Voronoi bins to generate the projected temperature map.

Figure~\ref{fig:macsj1752_thermo} shows the projected X-ray temperature map (see Figure~\ref{fig:macsj1752_temp_err} for fractional temperature uncertainty) of MACSJ1752, overlaid with X-ray and radio contours.
The fractional temperature uncertainty (Figure~\ref{fig:macsj1752_temp_err}) remains $\lesssim10\%$ within the subcluster regions and increases to $\sim$30\% near the radio relics.
The temperature structure is broadly elongated along the NE--SW direction, similar to the hypothesized merger axis. 
The SW and NE X-ray SB peaks have temperatures of $3.6\pm0.4$ and $5.0\pm0.7$~keV, respectively, and are statistically consistent with $1.7\sigma$ (Figure~\ref{fig:macsj1752_thermo}).

We further investigate the thermodynamic properties of MACSJ1752 by constructing electron-density, pressure, and entropy maps. 
For each Voronoi bin, we use the best-fit temperature and APEC normalization obtained from the spatially resolved spectral analysis.
The APEC normalization is defined as
\begin{equation}
{\cal N}_{\rm APEC} =
\frac{10^{-14}}{4\pi [D_{\rm A}(1+z)]^2}
\int n_{\rm e}n_{\rm H}\,dV,
\end{equation}
where $D_{\rm A}$ is the angular-diameter distance and $n_{\rm e}$ and
$n_{\rm H}$ are the electron and hydrogen number densities, respectively.
Assuming a uniform gas density within each bin, a volume-filling factor of
unity, and $n_{\rm e}=1.2n_{\rm H}$, we calculated the electron density as
\begin{equation}
n_{\rm e} =
\left[
\frac{1.2 \cdot 4\pi[D_{\rm A}(1+z)]^2\,10^{14}{\cal N}_{\rm APEC}}{V}
\right]^{1/2}.
\end{equation}
Following the geometric prescription of \citet{Lovisari2024}, the emitting volume of each bin was approximated as
\begin{equation}
V_i = 2A_i\sqrt{R_{500}^{2}-R_{{\rm proj},i}^{2}},
\end{equation}
where $A_i$ is the exact projected area of the irregular Voronoi bin and $R_{{\rm proj},i}$ is the projected distance of its centroid from the Planck SZ center \citep[$\alpha=268.0130^{\circ}, \delta=44.6695^{\circ}$;][]{Planck2016}.
We adopted the Planck-derived $R_{500c}$ (1.178~Mpc) for MACSJ1752. 
The pseudo-pressure and pseudo-entropy maps were then calculated as $P_{\rm e}=n_{\rm e}kT$ and $K=kTn_{\rm e}^{-2/3}$, respectively. 
Because this prescription implicitly assumes spherical symmetry, we tested the sensitivity to this assumption by adopting a constant line-of-sight depth of 1~Mpc.
We found that our qualitative interpretation of the thermodynamic structure was unchanged.

The electron-density map (upper right panel of Figure~\ref{fig:macsj1752_thermo}) presents two prominent density peaks associated with the two X-ray cores, together with elongated enhancements extending along the X-ray tails. 
The pressure distribution broadly follows the dense X-ray-emitting gas, with pressure maxima at the NE subcluster and WL clump, with elevated pressure extending into the tails.
The entropy map shows that the lowest-entropy gas ($K_{N,\rm min}=390$ and $K_{S,\rm min}=280$) remains concentrated around the two subclusters, whereas the gas extending into the tails generally has higher entropy ($K_{\rm tail}\approx400-500$) than the core minima. 
The persistence of relatively low-entropy gas at the cores suggests that the cores, particularly the southern subcluster, survived the high-speed collision, reminiscent of the ``Bullet'' \citep[e.g.,][]{2002markevitch, Markevitch2006, 2007markevitch}.
However, we do not find a significant dissociation of the ICM from the dark matter.

In combination, the compact low-entropy cores and the elongated, ram-pressure-stripped ICM tails are consistent with an off-axis merger with a non-zero impact parameter \citep[e.g.,][]{RichardLaferriere2023}.
In such a geometry, the dense central gas of the two subclusters may avoid a head-on collision, allowing the remnant cores to survive the first pericenter passage, while their outer gas is displaced and stripped by ram pressure. 
The relatively high density and pressure of the tails suggest that they contain gas removed from the progenitor subclusters, whereas their higher entropy relative to the core minima may reflect both the preferential stripping of high-entropy gas from larger radii and subsequent heating and partial mixing with the ambient ICM \citep[e.g.,][]{2011zuhone, Valdarnini2021}.

\subsubsection{ZWCL1856}

To construct the temperature map, we first defined a spatial aperture using a smoothed ObsID-merged 0.5--7.0~keV flux map. 
The aperture boundary was set by a contour corresponding to twice the median background level measured from the equivalently smoothed, merged blank-sky flux map. 
This restriction ensured that the spectral analysis was limited to regions with sufficiently significant source emission.

\begin{figure*}
    \centering
    \includegraphics[width=1\linewidth]{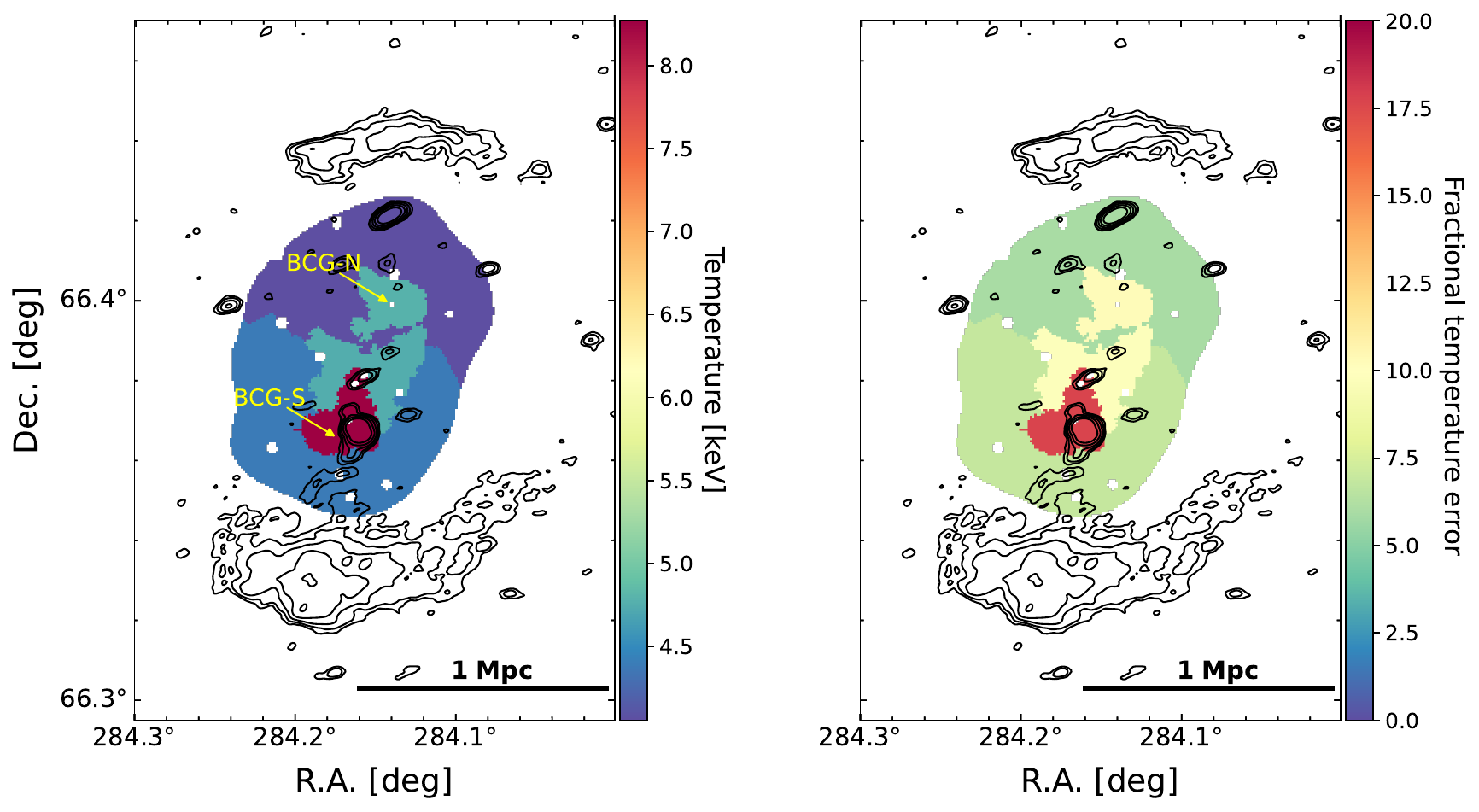}
    \caption{Projected temperature (left) and mean fractional temperature uncertainty (right) maps of ZWCL1856.
             Black contours trace the LOFAR radio emission, beginning at $3\sigma$ and increasing by factors of two, where $\sigma=127~\mu{\rm Jy~beam^{-1}}$.
             Yellow arrows in the left panel mark the northern (N) and southern (S) BCGs.
             The southern subcluster shows a substantially higher temperature than the northern subcluster.
            }
    \label{fig:zwcl_temp_map}
\end{figure*}

Within this aperture, we applied the \texttt{contbin} contour-binning algorithm \citep{Sanders2006} to the ObsID-merged broad-band counts image, together with the corresponding merged blank-sky background image. 
The algorithm defined spatial regions containing approximately 1,000 net source counts. 
Spectra for each contour-binned region were extracted and fitted following the procedure described above. 
The best-fit temperature in each region was then assigned to all corresponding pixels to produce the final temperature map.

We construct a temperature map to investigate the thermodynamic state of the ongoing merger in ZWCL1856. 
The cluster is generally cooler than MACSJ1752, consistent with its lower mass. 
Owing to the limited photon statistics in the cluster outskirts, we are unable to obtain reliable temperature measurements across the radio relics.

Based on the weak-lensing-calibrated mass--temperature relation of \citet{Mantz2016}, the WL masses measured by \citet{Finner2021} correspond to expected core-excised temperatures of $kT=2.0^{+0.6}_{-0.6}$~keV\footnote{The uncertainty includes the WL mass uncertainty and the intrinsic scatter, $\sigma_{\ln T}\simeq0.16$, of the mass--temperature relation.} and $kT=1.7^{+0.5}_{-1.0}$~keV for the northern and southern subclusters, respectively.
Applying the same relation to the total cluster mass of $2.4^{+0.9}_{-0.7}\times10^{14}~M_{\odot}$ yields an expected temperature of $kT=3.0^{+0.9}_{-0.7}$~keV for the system. 
By comparison, the temperature map shows values of $\sim$4~keV across most of the cluster outside the core regions. 
These temperatures are somewhat higher than, but statistically consistent with, the scaling-relation expectation.

One notable feature is the high core temperature of the southern subcluster, $8.3^{+1.7}_{-1.3}$~keV.
Although the two subclusters have similar WL masses, surrounding gas temperatures, and merger geometries, the southern subcluster is significantly hotter than the northern subcluster, which has a temperature of $4.8\pm0.5$~keV.
This corresponds to a temperature difference of $2.5\sigma$ between the northern and southern subclusters.

Although the southern subcluster has the largest fractional temperature uncertainty ($\sim$17\%) in the map, several tests suggest that the high temperature is unlikely to be caused solely by statistical noise. 
The corresponding region contains $\sim$1300 source counts across the 11 observations, the lowest count total among the temperature bins, which likely contributes to its comparatively large uncertainty. 
Nevertheless, the fitted spectrum shows no obvious anomalies, and the temperature posterior is well behaved. 
Moreover, a weaker enhancement persists in maps constructed using a different signal-to-noise threshold of 20 (Figure~\ref{fig:zwcl1856_temp_sn20}). 
Independent measurements using circular apertures also indicate a high temperature for the southern subcluster.
These consistency checks suggest that the southern ICM is hotter than the northern.
However, its precise temperature remains relatively uncertain because of the limited photon statistics.
In addition, cross-calibration studies have shown that {\it Chandra} can yield systematically higher temperatures than XMM-{\it Newton} and {\it NuSTAR}, particularly at high plasma temperatures, introducing a possible systematic uncertainty in the absolute temperature measurement \citep{Nevalainen2010, Schellenberger2015, Wallbank2022, Potter2023}.
Accordingly, although the relative temperature enhancement appears robust within the \textit{Chandra} data, its absolute value should be interpreted with caution.

\begin{figure*}
    \centering
    \includegraphics[width=1\linewidth]{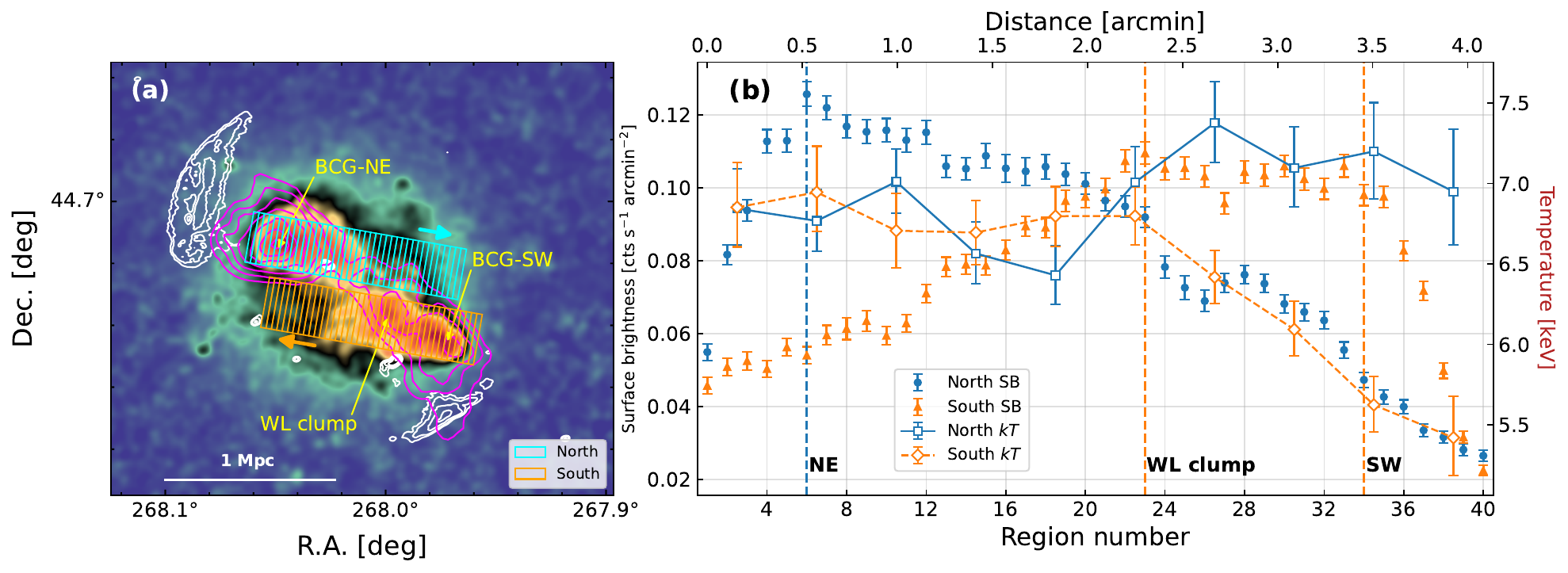}
    \caption{X-ray SB and projected-temperature profiles along the tails of the NE and SW subclusters in MACSJ1752.
             Left: XMM-{\it Newton} 0.5--2.0~keV image with the profile extraction regions overlaid.
             The cyan and orange boxes correspond to the NE and SW subclusters, respectively, with the colored arrows indicating the profile directions from each subcluster core toward its trailing X-ray tail.
             The BCG positions are indicated in yellow.
             White contours show the LOFAR 144~MHz radio emission, beginning at $3\sigma$, where $\sigma=280~\mu\mathrm{Jy~beam^{-1}}$, and increasing by factors of two.
             Magenta contours show weak-lensing mass-map signal-to-noise levels from 5 to 11.
             Right: Background-subtracted X-ray SB profiles for the NE (blue circles) and SW (orange triangles) tails, together with their projected-temperature profiles.
             The NE temperatures are shown by blue open squares connected by a solid line, and the SW temperatures by orange open diamonds connected by a dashed line.
             Each temperature bin combines four consecutive SB regions.
             The vertical dashed lines mark the regions containing BCG-NE, the WL clump, and BCG-SW.
             The upper axis gives the projected distance from region 1.
            }
    \label{fig:macsj1752_tail}
\end{figure*}

\subsection{X-ray tails of the subclusters}
Both of these clusters exhibit elevated X-ray emission between the subclusters. 
These tail-like features are expected to form in post-collision mergers and evolve with time.

\subsubsection{MACSJ1752}

The soft-band X-ray morphology of MACSJ1752 is characterized by exceptionally long, Mpc-scale tail-like structures aligned with the NE--SW merger axis. 
To characterize these structures and investigate their origins, we present the SB and projected-temperature profiles of both subclusters in Figure~\ref{fig:macsj1752_tail}.

To quantify the physical properties of these structures, we extracted background-subtracted 0.5--2.0~keV SB profiles through the rectangular regions shown in Figure~\ref{fig:macsj1752_tail}(a), after masking point sources. 
Each profile contains 40 adjacent SB bins, each $\sim$32~kpc wide at the cluster redshift, and spans $\sim$1.2~Mpc. 
The NE regions are numbered from the outer side through the BCG near region 6 toward the intercluster region, whereas the SW regions are numbered in reverse order, with the intercluster side at low region numbers and the BCG near region 34.
Temperatures were measured in ten coarser spatial bins, each formed by combining four consecutive SB bins.

Figure~\ref{fig:macsj1752_tail}(b) shows that the SB profiles of both the NE and SW subclusters peak at their respective BCGs and are markedly asymmetric about both peaks.
On the outer, leading sides of the subclusters, the X-ray SB decreases rapidly, whereas extended emission trails from each core toward the center of the system. 
The oppositely directed tails overlap in projection and form a continuous bridge of X-ray-emitting gas between the subclusters. 
The NE component has the more sharply peaked X-ray core and a comparatively narrow, slightly curved wake. 
Meanwhile, the SW component has a more bullet-like head followed by broader emission that remains prominent over a projected extent of $\sim$350 kpc. 
This plateau extends slightly beyond the WL clump\footnote{This feature is labeled ``center'' in \citet{Finner2021}.} and galaxy overdensity identified by \citet{Finner2021}, after which the SB begins to decline.
The spatial coincidence of the SB plateau with the WL clump and galaxy overdensity suggests that the extended X-ray feature may not be produced solely by gas stripped from the SW subcluster. 
Instead, it may represent a superposition of stripped gas and emission associated with an additional lower-mass subcluster.
We discuss this interpretation further in Section~\ref{subsec:sim}.

Figure~\ref{fig:macsj1752_tail} also reveals distinct temperature distributions between the two subclusters. 
The temperature profile of the NE subcluster remains approximately constant over most of the sampled range, with a modest increase toward larger distances. 
At still larger radii, the temperature map in Figure~\ref{fig:macsj1752_thermo} shows an enhancement along the putative shock front extending inward from the SE relic. 
Motivated by this feature, we searched for a corresponding X-ray shock signature using the SB and temperature profiles presented in Appendix~\ref{app:macsj1752_west}. 
Although the temperature decreases across the putative front, no statistically significant SB discontinuity is detected. 
Thus, while the temperature structure is suggestive of shock heating, the current X-ray data do not provide sufficient evidence to confirm the presence of a shock at this location.

The SW subcluster exhibits a clear cool-core signature, with a central temperature substantially below the cluster average \citep[$\sim$7~keV;][]{Finner2021}.
Beyond the core, the temperature rises to $\sim$6.5~keV near the WL clump and remains approximately constant farther along the profile, indicating a transition to hotter, nearly isothermal gas.

\begin{figure*}
    \centering
    \includegraphics[width=1\linewidth]{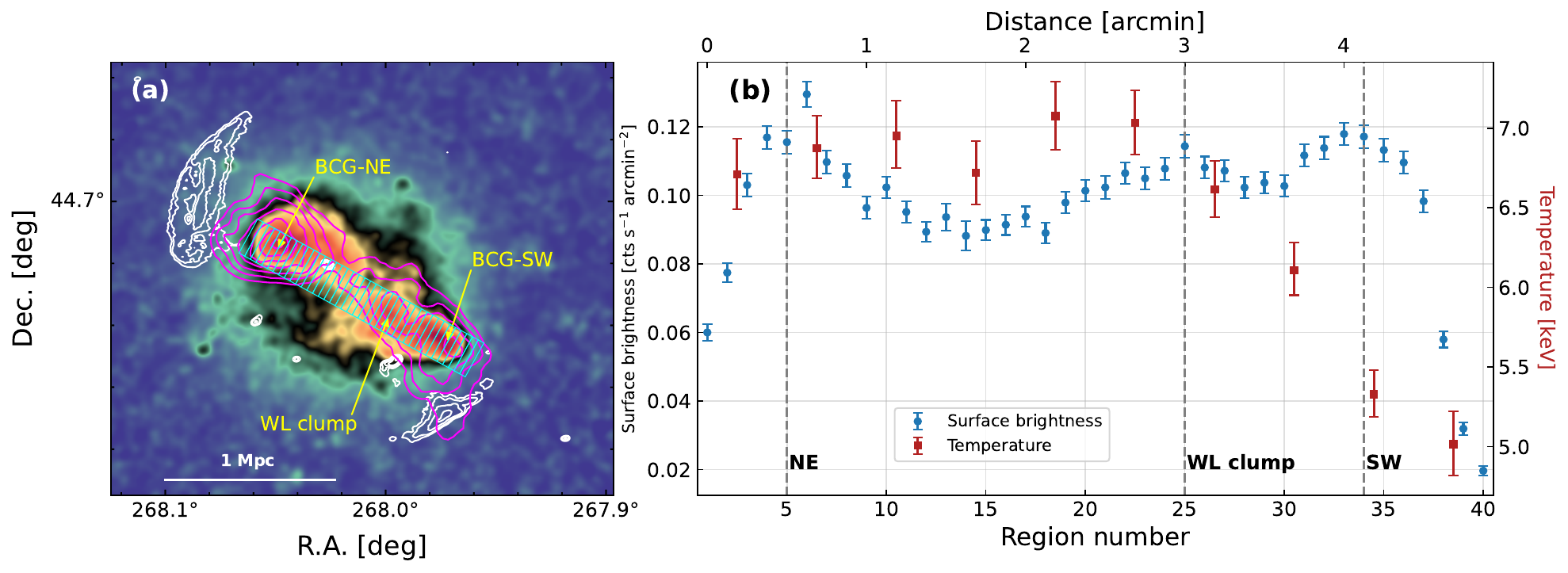}
    \caption{X-ray SB and projected-temperature profiles across the two subclusters of MACSJ1752 along the NE--SW merger axis. 
             Left: XMM-{\it Newton} 0.5--2.0~keV image with the cyan profile extraction regions overlaid. 
             The positions of BCG-NE and BCG-SW are indicated in yellow. 
             White contours show the LOFAR 144~MHz radio emission, beginning at $3\sigma=840~\mu\mathrm{Jy~beam^{-1}}$ and increasing by factors of two, while magenta contours show weak-lensing mass-map signal-to-noise levels from 5 to 11. 
             Right: Background-subtracted X-ray SB (blue circles; left axis) and projected temperature (red squares; right axis) measured along the same regions. 
             The region number increases from the NE side toward the SW side of the system, and the upper axis gives the corresponding projected distance from region 1.
             The vertical dashed lines mark the projected positions of the two BCGs.
             Each temperature measurement was obtained by combining four consecutive SB regions.
             }
    \label{fig:macsj1752_tail2}
\end{figure*}

Figure~\ref{fig:macsj1752_tail2} provides a complementary profile through both subclusters along the full merger axis. 
Two SB maxima are detected near the projected BCG positions.
Between the two peaks, the SB remains roughly constant, showing that diffuse X-ray emission fills the inter-core region. 
Outside the two BCGs, the SB falls rapidly, consistent with the compact leading edges seen in the X-ray SB image.
We discuss the origin of the nearly constant SB along the X-ray tail through comparison with simulated analogs in Section~\ref{subsec:sim}.

The gas between the subclusters is also the hottest component sampled by the axial profile. 
Between the BCGs, the projected temperature is $kT\sim7$~keV, with a mild increase near the middle of the system. 
Toward the SW subcluster and its outer side, the temperature declines rapidly, which is already presented in Figure~\ref{fig:macsj1752_tail}.
The combination of enhanced temperature and sustained SB between the cores is consistent with compression and heating of the ICM during the merger, while the cooler gas around the SW peak likely traces the surviving remnant of the SW subcluster core. 
Together, the two profiles show that both gaseous cores remain intact, but their downstream structures and thermal evolution are substantially different.

\subsubsection{ZWCL1856} \label{subsubsec:zwcl1856_tail}

\begin{figure*}
    \centering
    \includegraphics[width=1\linewidth]{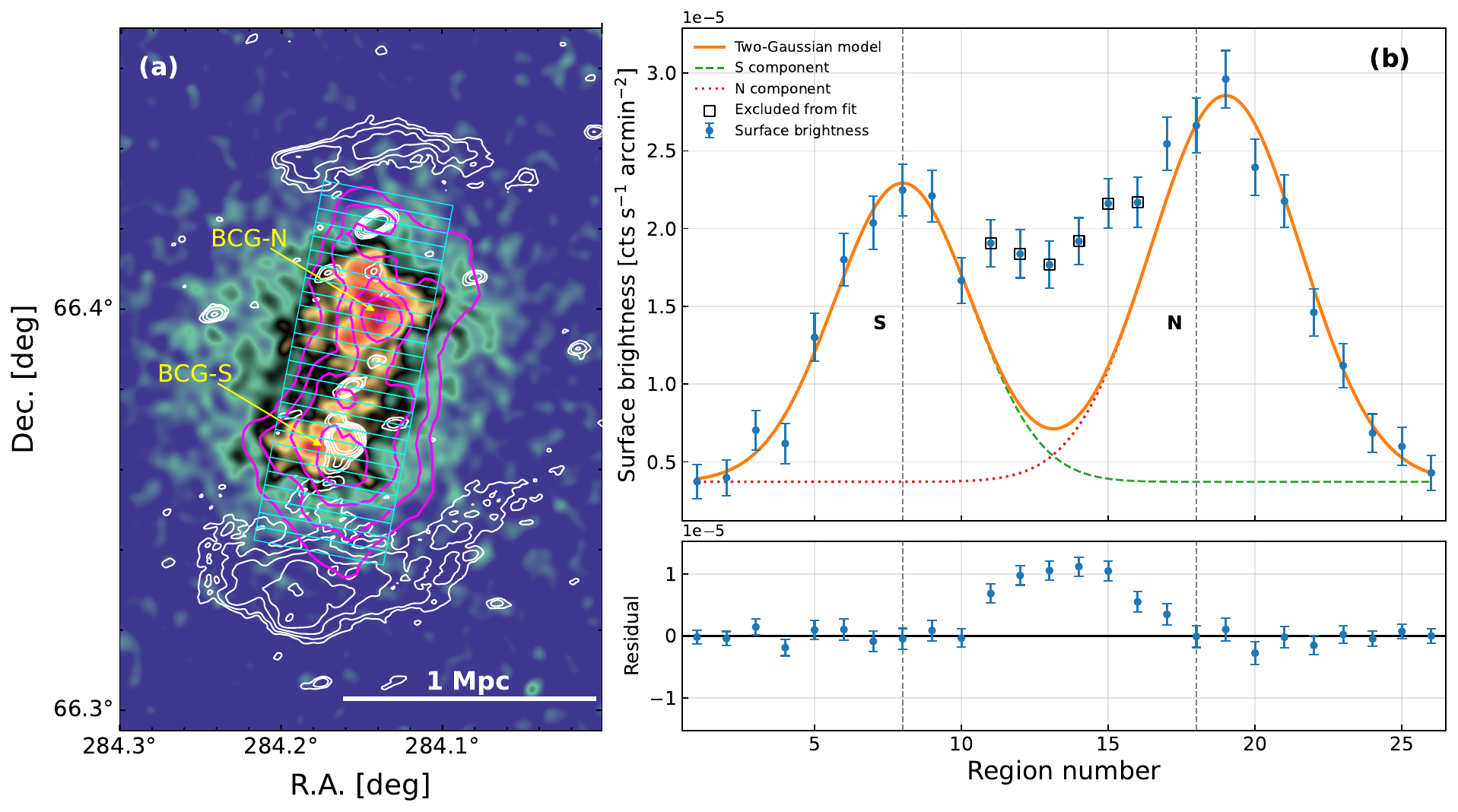}
    \caption{{\it Chandra} X-ray image and SB profile of ZWCL1856.
             Left: Gaussian-smoothed ($\sigma=4\farcs5$) X-ray image overlaid with LOFAR radio contours in white and WL signal-to-noise contours in magenta.
             The cyan boxes indicate the regions used to extract the SB profile, while the yellow arrows mark the southern (S) and northern (N) BCGs.
             Top right: X-ray SB profile measured in consecutive extraction regions.
             The solid curve shows the sum of two Gaussian components and a constant background, with the Gaussian centers fixed at the X-ray peaks (regions 8 and 19).
             The projected positions of the southern and northern BCGs are marked by the vertical dashed lines.
             The dashed and dotted curves show the southern and northern Gaussian components plus a constant background, respectively.
             Open squares mark the regions excluded from the fit.
             The spatial temperature structure of ZWCL1856 is examined independently in Figure~\ref{fig:zwcl_temp_map}.
             Bottom right: Residual SB profile obtained by subtracting the best-fit two-Gaussian model from the observed profile.
             This residual highlights excess X-ray emission between the two subclusters, which may trace merger-displaced gas.
            }
    \label{fig:zwcl_tail}
\end{figure*}

The X-ray SB map of ZWCL1856 shows prominent peaks offset from their respective BCGs, together with extended, clumpy emission filling the intercluster region.
To quantify this emission, we measure the SB profile along the putative merger axis passing through the two X-ray peaks, as shown in Figure~\ref{fig:zwcl_tail}, while masking the identified point sources (Section~\ref{sec:observations}).
The profile exhibits two prominent peaks near the projected positions of the southern and northern BCGs, with the northern subcluster being the brighter peak.
The SB remains elevated throughout the region between them.

To isolate the tail-like emission from the components associated with the core regions of the individual subclusters, we fit the SB profile with two Gaussian functions plus a constant background.
The Gaussian centroids are fixed at the X-ray peaks of the southern (S) and northern (N) subclusters (regions 8 and 19, respectively), while regions 11--16 are excluded from the fit to prevent the excess intercluster emission from biasing the model.
The residual tail profile reveals a pronounced, asymmetric excess of X-ray emission between the two subclusters.
This excess rises abruptly at its interface with the southern subcluster but connects more gradually to the northern subcluster. 
This asymmetry suggests that this excess emission may be more closely associated with gas displaced from the disturbed northern subcluster.

\section{Discussion} \label{sec:discussion}
Deep X-ray observations of MACSJ1752 and ZWCL1856 provide new context for understanding these merging systems. 
The increased depth allows us to spatially resolve the thermodynamic properties of the clusters that are critical to understanding their merger histories.

\subsection{Updated merger scenario} \label{subsec:merger_scenario}

Despite their similar radio relic and WL configurations, MACSJ1752 and ZWCL1856 appear to represent different stages of cluster-merger evolution. 
The larger relic radii in ZWCL1856 relative to its characteristic cluster radius are consistent with a later merger stage, in agreement with the X-ray morphology.
The deep X-ray observations reveal distinct relationships among the ICM, galaxies, and dark matter that allow us to refine the merger scenario for each system.

In \textbf{MACSJ1752}, the X-ray, WL, and galaxy distributions consistently identify two equal-mass subclusters located to the NE and SW \citep{Finner2021}.
The high-speed collision has displaced some of the ICM gas from the dark matter potential \citep[e.g.,][]{Markevitch2006, 2019sheardown}, but no significant dissociation is detected.
The curved tail of the NE subcluster extends toward the intercluster region, suggesting transverse gas motions associated with an off-axis merger.
On the opposite side of the system, in X-rays, the SW subcluster exhibits a bullet-like morphology, consisting of a compact, low-entropy core bounded by a prominent cold front and followed by an extended trailing tail.
Unlike the Bullet cluster, however, the dense core of MACSJ1752 has not become significantly separated from its associated dark matter peak \citep[e.g.,][]{Markevitch2006, 2006clowe, 2025cha_bullet, Cho2026, Rihtarsic2026}.

The X-ray emitting gas density and temperature discontinuities detected at the NE relic provide direct evidence that an outward-propagating merger shock has reached the cluster outskirts.
Together with the surviving cores and the Mpc-scale trailing gas tails, this configuration is consistent with a post-merger system viewed shortly after first core passage.
A non-zero impact parameter may have allowed the dense cores to survive the destruction of the cool core while stripping leaves an extended X-ray tail.

\textbf{ZWCL1856} exhibits a different hot gas configuration.
The southern X-ray peak is spatially offset to the east of its BCG and the corresponding WL mass peak \citep{Finner2021}. 
The northern X-ray peak is displaced toward the outer, relic-facing side of the system relative to the northern BCG and WL peak. The elevated X-ray emission between the subclusters and the existence of radio relics indicate that ZWCL1856 is a post-collision merging system.
The observed configuration, in which the gas lies ahead of the BCG and mass peak along the inferred direction of motion, is therefore difficult to reconcile with the scenario described for MACSJ1752.

A more natural explanation is that ZWCL1856 is well-past first pericenter passage and is currently near or shortly after its first apocenter crossing. 
As the subcluster approaches apocenter, its velocity and the associated ram pressure decrease, allowing gas displaced during core passage to expand and overtake the decelerating dark matter potential. 
This process can produce a slingshot tail or an apparent leading X-ray feature \citep{2019sheardown}.
A comparable evolution was demonstrated for the NGC~4839 group in the Coma cluster \citep[e.g.,][]{Lyskova2019, Churazov2021, HyeongHan2026arXiv}.
Although the NGC~4839 group X-ray morphology initially suggested first infall, hydrodynamical simulations and the observed gas fraction favored a post-merger scenario in which the displaced gas fell back toward, and overshot, the group potential near apocenter.
El Gordo provides another relevant comparison.
Its southeastern X-ray peak lies farther from the cluster center than the corresponding dark matter peak \citep[e.g.,][]{Menanteau2012, Jee2014, Kim2021}, a configuration interpreted as a returning, post-apocenter merger phase \citep{Ng2015}.

We therefore interpret MACSJ1752 as an off-axis merger observed during an active, outbound post-pericenter phase, when gas stripping remains prominent, whereas ZWCL1856 is likely observed at a more advanced stage near or after first apocenter, when slingshot-like gas motions can develop. 
This comparison demonstrates that similar double-relic separations and dark matter configurations do not necessarily imply identical merger stages. 
The observable ICM morphology is also shaped by the impact parameter, progenitor core structure, and previous assembly history.

\subsection{Analogs in cosmological simulations} \label{subsec:sim}

Our deep X-ray observations identified unique features that are valuable for diagnosing ongoing dynamical activity in both cluster mergers.
In MACSJ1752, the SB and temperature discontinuities detected at the NE relic provide X-ray evidence for a merger-driven shock (Section~\ref{subsec:macsj1752_shock}), complementing the evidence obtained from radio observations \citep{vanWeeren2012, 2012bonafede, Chiesa2026arXiv}. 
The spatial analysis in MACSJ1752 further shows an extended X-ray tail between the colliding clusters, along which the SB varies by only $\sim$20\% over much of its extent. 
The origin of such a long and nearly uniform X-ray structure is difficult to determine from the observations alone.
We therefore search for morphologically similar systems in cosmological simulations to investigate the merger histories capable of reproducing these features.

\begin{figure}
    \centering
    \includegraphics[width=0.9\linewidth]{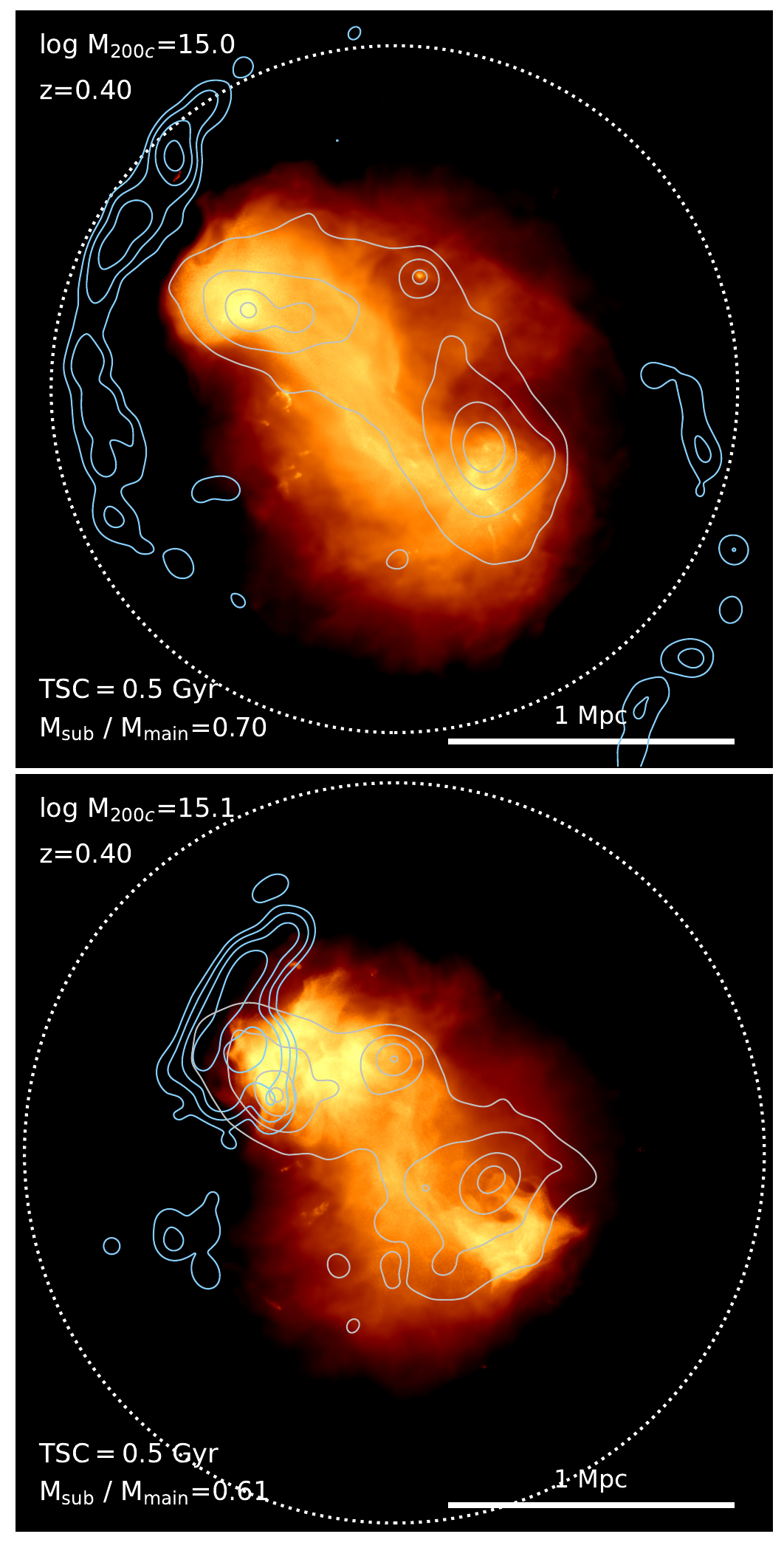}
    \caption{Projected X-ray SB maps of two morphological analogs of MACSJ1752 and ZWCL1856 identified in TNG-Cluster at $z=0.4$.
             The projections have been rotated to match the orientation of MACSJ1752.
             Gray and blue contours show the projected dark matter distribution and radio emission from merger shocks at 1.4~GHz, respectively.
             The circle marks $R_{\rm 500c}$, and the X-ray color scale spans two orders of magnitude to display both the bright cores and extended low SB structures.
             We note that the masses of the analogs are an order of magnitude higher than ZWCL1856.
            }
    \label{fig:tng_tail}
\end{figure}

We use the TNG-Cluster suite of cosmological magnetohydrodynamical zoom-in simulations \citep{Nelson2024} to find the analogs.
TNG-Cluster follows the formation of massive galaxy clusters selected from a Gpc-scale parent volume and incorporates a comprehensive galaxy formation model, the cosmological evolution of magnetic fields, and the identification of shock waves \citep{Lee2024}.
Its combination of a cosmological environment and high spatial resolution allows us to examine whether extended X-ray tails arise naturally during cluster mergers and whether they are associated with more complex assembly histories.

To identify cluster mergers with MACSJ1752-like bright, extended X-ray tails, we construct projected X-ray SB maps of the simulated clusters at $z=0.4$. 
We calculate the X-ray emissivity of each gas cell in the energy range of 0.5--2.0~keV in the observer frame (redshifted) by using the APEC model and the properties of the gas cell.
For each projected map, we measure the average X-ray SB within 50~kpc of the main cluster center and define the tail extent as the maximum distance reached by the contour corresponding to one-half of this central value. 
Among 352 simulated clusters, 62 systems have X-ray extensions exceeding 500~kpc, although many of these arise from irregular, disturbed ICM morphologies in massive, unrelaxed clusters rather than from coherent merger-driven tails.
We therefore visually inspect these systems and identify two clusters with distinct X-ray cores accompanied by extended X-ray tails.

For comparison with the observed multiwavelength features, we also construct projected dark matter and diffuse radio maps of the simulated systems.
The projected dark matter mass map is generated by smoothing the dark matter particle distribution with a Gaussian kernel of width equal to half the radius enclosing 64 neighboring particles. 
The diffuse radio emission from merger shocks at 1.4~GHz is overlaid on the X-ray map and modeled using the magnetic-field strength and the nonthermal plasma content. 
The model assumes that a Mach-number-dependent fraction of the shock-dissipated energy is converted into nonthermal energy \citep[e.g.,][]{Hoeft2007, Lee2024}.
Following \citet{Lee2026}, we exclude shocks associated with individual galaxies and large-scale accretion when generating the radio maps.

Figure~\ref{fig:tng_tail} presents two X-ray analogs at $z=0.4$, both with a time since collision (TSC) of 0.5~Gyr.
As in MACSJ1752 and ZWCL1856, both clusters contain extended X-ray-emitting structures connecting the two merging subclusters. 
The pericenter distance was $\sim$300~kpc, corresponding to a merger with a non-zero impact parameter.
This geometry may have facilitated the survival of the dense X-ray cores while allowing sufficient ram pressure to strip and compress the outer gas, thereby producing the extended X-ray tails.
Most of the displaced gas remains between or trails the dark matter peaks, although some gas begins to move ahead of the corresponding mass peak, marking the development of a possible slingshot tail \citep[e.g.,][]{2019sheardown, Kim2026}.

Although the TNG-Cluster analogs reproduce the extended X-ray tail morphology, the radio relics in the simulated analogs are substantially more asymmetric than those observed in MACSJ1752 and ZWCL1856. 
The analogs are also considerably more massive than ZWCL1856, implying higher characteristic sound speeds that can affect the shock propagation velocities and relic locations.
In both analogs, the radio relic ahead of the eastern X-ray core remains relatively close to the X-ray core, with a separation broadly comparable to that of the NE relic and the associated X-ray shock in MACSJ1752.
The double relic counterpart in the top panel of Figure~\ref{fig:tng_tail} lies at a much larger clustocentric distance than the SW relic of MACSJ1752 and is detached from the X-ray core. 
In the bottom panel of Figure~\ref{fig:tng_tail}, the corresponding shock is located at a comparable distance, but its radio luminosity is more than two orders of magnitude lower.
The displayed analogs should therefore be regarded primarily as X-ray morphological analogs rather than exact counterparts of the observed double-radio-relic clusters.

\begin{figure}
    \centering
    \includegraphics[width=\linewidth]{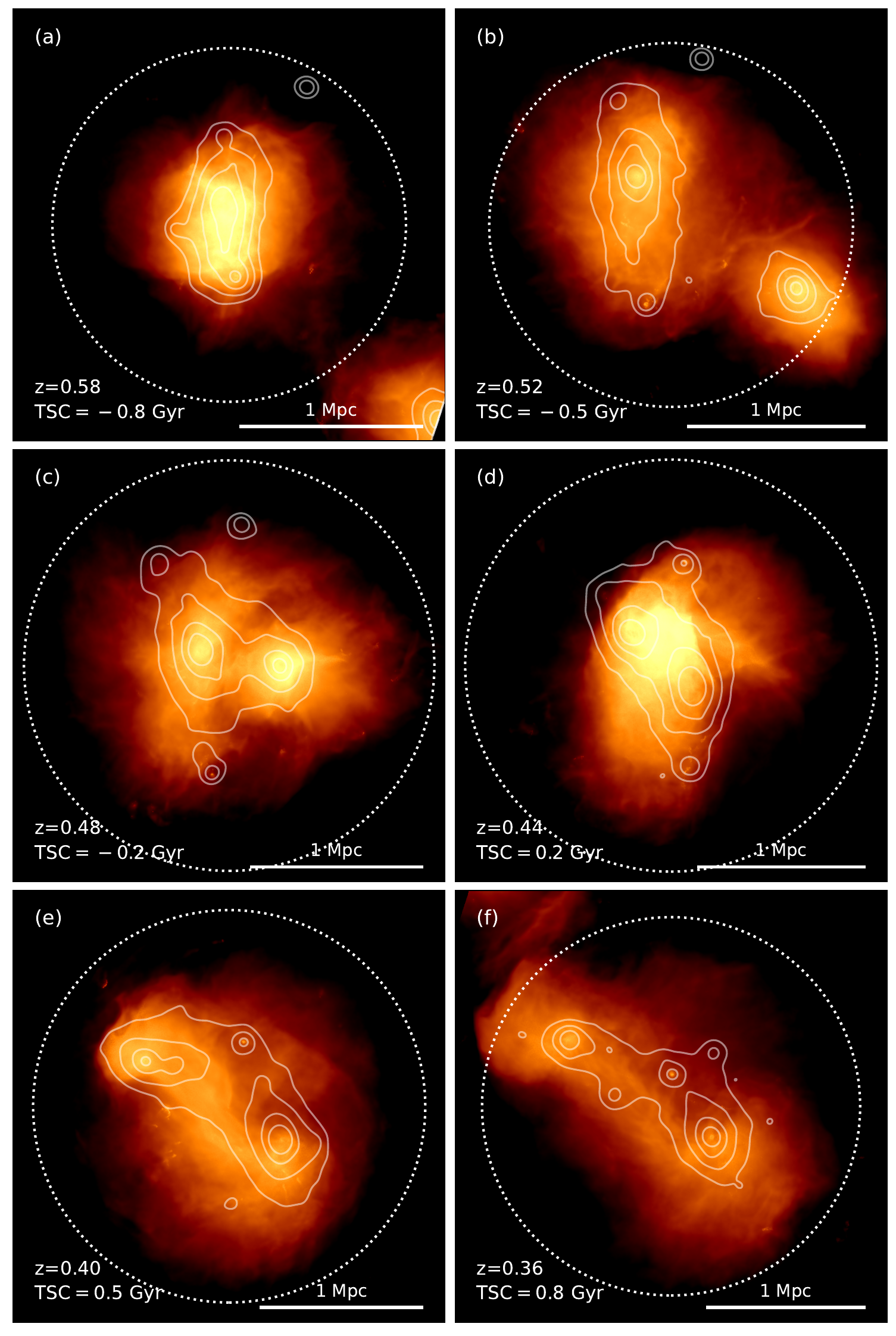}
    \caption{Evolution of the projected X-ray SB distribution of a representative TNG-Cluster analog through different merger phases.
             White contours show the smoothed projected mass distribution, and the dotted circles mark $R_{500c}$ at each snapshot.
            }
    \label{fig:tng_evol}
\end{figure}

Extended X-ray tails with nearly uniform X-ray SB are uncommon among the off-axis mergers examined in our search in the TNG-Cluster simulations.
Many mergers instead show a pronounced decline in SB with increasing distance from the X-ray core.
A notable shared characteristic of the two selected analogs is that the primary cluster experienced a minor merger shortly before its major merger with the secondary cluster.

Figure~\ref{fig:tng_evol} illustrates this sequence for one representative system (top panel of Figure~\ref{fig:tng_tail}).
Before the major merger (Figure~\ref{fig:tng_evol}(a)), a tertiary halo with a mass ratio of $\sim$0.1 merged with the primary cluster and disturbed its ICM, producing a more extended and elliptical gas distribution.
Approximately 0.5~Gyr later (Figure~\ref{fig:tng_evol}(b)--(d)), the secondary cluster encountered this pre-disturbed ICM during the pericenter passage.
The interaction stripped and compressed the extended gas, producing a long, bright X-ray structure between the two surviving cores (Figure~\ref{fig:tng_evol}(e)).

At a later merger stage (Figure~\ref{fig:tng_evol}(f)), some of the displaced gas overtakes the associated dark matter potential and begins to extend ahead of the mass peak.
This evolution produces a slingshot-like morphology qualitatively similar to that observed in ZWCL1856, where bright X-ray emission extends laterally and beyond the BCGs (Figure~\ref{fig:zwcl_tail}).
In both the simulated analog and ZWCL1856, the X-ray SB of the intracluster gas also declines with increasing distance from the subcluster cores.

The shared merger histories of the selected analogs provide insight for the unusual X-ray morphology of MACSJ1752.
In particular, the third mass clump identified in its weak-lensing analysis \citep{Finner2021} may represent evidence of an additional substructure involved in the merger.
MACSJ1752 may therefore have experienced a preceding minor merger before the current major collision, rather than evolving through an isolated binary merger.
The late phase of this analog (Figure~\ref{fig:tng_evol}), during which the gas begins to overrun the dark matter potentials and form slingshot-like structures, may instead resemble the more advanced merger phase inferred for ZWCL1856.
Nevertheless, these analogs do not uniquely determine the histories of the observed clusters.
They demonstrate that a preceding minor merger followed by an off-axis major collision provides one physically plausible pathway for producing the observed extended X-ray structures.

\subsection{Radio relic clusters as probes of self-interacting dark matter}

Understanding the properties of dark matter is a central question in astrophysics. 
Merging galaxy clusters have been used as a probe of dark matter, for example to test self-interacting dark matter (SIDM) models \citep{2004markevitch, 2008randall, 2015harvey, 2025cha_bullet, Rihtarsic2026}. 
They provide unique constraints in the high-velocity regime where cluster collisions probe the SIDM cross section at velocities of $\sim$1,000--3,000~km~s$^{-1}$ \citep[e.g.,][]{Bechtol2022arXiv}.
However, not all merging clusters present the key features needed to constrain dark matter properties.
Dissociative mergers that are taking place in the plane of the sky are ideal because they present the largest separations of the ICM from the dark matter potential. 
Clusters hosting double radio relics may also be valuable for constraining SIDM because the radio relics trace merger-induced shocks and constrain the chronology of the event. 
Both MACSJ1752 and ZWCL1856 are initially appear to be promising candidates candidates for dark matter studies.

Recently, \citet{Jee2026} used double-radio-relic clusters to put the most stringent constraint on the SIDM cross section to date ($\sigma/m<0.22$ cm$^{2}$ g$^{-1}$). 
Their method uses the separation between the two radio-relic shocks as a merger chronometer and compares it with the halo-to-halo separation measured from weak lensing.
Our multiwavelength analyses of MACSJ1752 and ZWCL1856 present an interesting juxtaposition in this context. 
The two clusters exhibit similar, bimodal weak-lensing mass distributions with MACSJ1752 having a halo-to-halo separation of 1.2 Mpc and ZWCL1856 a separation of 600 kpc. 
Interestingly, both host double radio relics that are physically separated by $\sim$2~Mpc, and show merger geometries that would initially appear favorable for SIDM studies. 
However, the realization that MACSJ1752 is observed shortly after collision and ZWCL1856 is near apocenter indicates that MACSJ1752 is a more valuable system for SIDM studies as follows.

MACSJ1752 shows a clear bimodal X-ray morphology and thermodynamic features consistent with an active, major merger after a first pericenter passage, making it well suited to the dynamical framework adopted for SIDM constraints. 
Indeed, \citet{Jee2026} find MACSJ1752 to be the most constraining individual system in their sample. 
In contrast, our X-ray analysis suggests that ZWCL1856 is likely near apocenter or has begun its return toward a second core passage. 
In such a late merger stage, the relation between the halo separation and TSC is no longer described by a simple outgoing first-passage trajectory, potentially introducing a degeneracy between merger phase and the halo deceleration expected from SIDM.
Thus, systems that appear similar based on their projected weak-lensing and radio morphologies can have substantially different utility for SIDM constraints.

This comparison highlights the importance of accurately determining the dynamical state of individual mergers before incorporating them into stacked SIDM analyses. 
Selection based solely on bimodal mass distributions and double-radio-relic morphology may not be sufficient to guarantee a simple merger geometry and evolutionary stage. 
Detailed multiwavelength analyses, particularly combining X-ray thermodynamic information with weak-lensing and radio observations \citep[e.g.,][]{Rajpurohit2024, Rajpurohit2025, HyeongHan2024_a746, Stroe2025, Ahn2025}, are therefore essential for identifying systems whose merger histories can be robustly interpreted in the context of SIDM.

\section{Conclusion} \label{sec:conclusion}

We have presented a comparative X-ray study of the double-radio-relic merging clusters MACSJ1752 and ZWCL1856 using deep XMM-{\it Newton} and {\it Chandra} observations, respectively. 
Although both systems are equal-mass mergers viewed close to the plane of the sky and have similar relic separations, their X-ray morphologies reveal substantially different ICM dynamics and merger stages. 
Our results are summarized as follows.

\begin{itemize}
    \item In MACSJ1752, we detect X-ray SB and temperature discontinuities spatially coincident with the NE radio relic. 
    The measured density compression, $C_{\rho}=1.6\pm0.3$, corresponds to $\mathcal{M}_{\rho}=1.4\pm0.2$. 
    The temperature jump yields $\mathcal{M}_{kT}=1.4\pm0.4_{\rm stat}\pm0.2_{\rm sys}$ with a $2\sigma$ upper limit of $\mathcal{M}_{kT}\lesssim2.3$.
    Both values are lower than the radio-derived Mach number of $\mathcal{M}_{\rm radio}\sim3$.
    This discrepancy is consistent with the radio and X-ray measurements probing different effective regions of an inhomogeneous shock surface. 
    At the SW relic, we find no coherent discontinuity in either SB or temperature across the relic. 
    The prominent inner SB edge instead traces the cold front associated with the surviving SW core. 

    \item The thermodynamic maps of MACSJ1752 reveal two compact, relatively low-entropy cores embedded in Mpc-scale stripped gas. 
    The X-ray SB remains nearly constant between the two subclusters, varying by only $\sim20\%$ over much of the inter-core region. 
    In particular, the SW tail exhibits an approximately 350-kpc-long SB plateau that extends slightly beyond the third WL mass concentration and associated galaxy overdensity. 
    This spatial correspondence suggests that the extended emission may contain both gas stripped from the SW subcluster and gas associated with an additional lower-mass subcluster. 
    Together, the surviving cores and asymmetric tails favor an off-axis collision with a non-zero impact parameter and possibly a more complex assembly history than an isolated binary merger.

    \item In ZWCL1856, we find no clear SB discontinuity at either radio relic, although the limited photon statistics do not exclude the presence of weak shocks. 
    In addition, reliable upstream and downstream temperature measurements cannot be obtained at the relic fronts. 
    The temperature map nevertheless reveals a pronounced asymmetry. 
    The southern subcluster has a temperature of $8.3^{+1.7}_{-1.3}$~keV, compared with $4.8\pm0.5$~keV for the northern subcluster, corresponding to a difference of $2.5\sigma$. 
    The northern X-ray peak is displaced toward the outer, relic-facing side relative to its BCG and WL peak, while the residual SB profile reveals asymmetric excess emission between the subclusters. 
    These features are consistent with ZWCL1856 being observed at a more advanced merger stage, near or after first apocenter, when previously displaced gas can overtake the associated dark matter peak and develop a slingshot-like morphology.

    \item We identify two X-ray morphological analogs in TNG-Cluster that contain surviving cores connected by extended X-ray-emitting structures. 
    Both analogs experienced a minor merger that disturbed and extended the primary ICM shortly before an off-axis major collision. 
    The subsequent major merger stripped and compressed this pre-disturbed gas, producing long, bright X-ray tails. 
    At later times, some of the displaced gas overtakes the corresponding dark matter peak and develops a slingshot-like morphology resembling that inferred for ZWCL1856. 
    This evolutionary sequence provides a plausible connection between the active trailing structures in MACSJ1752 and the laterally extended gas in ZWCL1856. 
    The third WL mass concentration in MACSJ1752 may represent evidence of a similarly complex assembly history, although the observations do not uniquely establish this scenario. 
    Moreover, the simulated systems should be regarded as X-ray morphological analogs rather than exact counterparts because their radio relic emission is substantially more asymmetric than observed.

\end{itemize}

Taken together, our results favor an off-axis, outbound post-pericenter configuration for MACSJ1752, in which ram-pressure stripping remains active, and a more advanced, near- or post-apocenter configuration for ZWCL1856, in which slingshot-like gas motions have begun to develop. 
Thus, similar double-relic separations and WL mass configurations do not necessarily imply equivalent ICM dynamics or merger stages. 
The observed gas distribution also depends on the impact parameter, progenitor core structure, and previous assembly history.

This difference in merger stage has important implications for using double-radio-relic clusters to constrain the properties of dark matter. 
Although no significant ICM--dark-matter dissociation is detected in MACSJ1752, its active outbound merger configuration makes the ordering of its components more straightforward to interpret. 
In contrast, the inferred near- or post-apocenter stage of ZWCL1856 complicates the interpretation of gas--mass offsets and may limit its utility for SIDM constraints.
Spatially resolved X-ray observations, combined with radio, WL, galaxy, and simulation analyses, are essential for reconstructing merger histories and identifying clusters for robust dark matter constraints.

\section*{Acknowledgments}

KH is supported in part by the Open Universe effort, which is funded by NASA under JPL Contract Task 70-711320, ``Maximizing Science Exploitation of Simulated Cosmological Survey Data Across Surveys'' and supported in part by NASA grant 22-ROMAN11-0011, ``Maximizing Cosmological Science with the Roman High Latitude Imaging Survey.'' 
Support for KF and KH was provided by the National Aeronautics and Space Administration through Chandra Award Number GO3-24109X issued by the Chandra X-ray Center, which is operated by the Smithsonian Astrophysical Observatory for and on behalf of the National Aeronautics Space Administration under contract NAS8-03060.
The work was partially supported by NASA under award number 80GSFC24M0006.
W.L. acknowledges the support of the National Research Foundation of Korea (NRF) grant funded by the Korea government (RS-2024-00340949).
A.T. acknowledges the support from NASA under award number 80GSFC24M0006.
Support for SWR and JAZ were provided by the Chandra X-ray Center through NASA contract NAS8-03060 and the Smithsonian Institution.
M.J.J. acknowledges support from the National Research Foundation of Korea (NRF) grants funded by the Korea government (MSIT), Nos. 2022R1A2C1003130 and RS-2026-25607498.
This work is based on observations obtained with XMM-Newton, an ESA science mission with instruments and contributions directly funded by ESA Member States and the US (NASA). 
This research has made use of observations made by the Chandra X-ray Observatory, and software provided by the Chandra X-ray Center (CXC).
We thank the TNG-Cluster team for allowing us to utilize the entire simulation data set, which helps us to reconstruct the merging scenario.
The calculations with TNG-Cluster in this paper were carried out on the Vera cluster of the Max Planck Institute for Astronomy (MPIA).

\appendix
\twocolumngrid

\renewcommand\thefigure{\thesection.\arabic{figure}}
\setcounter{figure}{0}

\renewcommand\thetable{\thesection.\arabic{table}}
\setcounter{table}{0}

\section{X-ray Observations} \label{app:obs}

Table~\ref{tab:xray_obs} presents the X-ray observations used in this study.
It includes the ObsIDs, instruments, pointing coordinates, observation dates, and cleaned exposure times.
For MACSJ1752, the total cleaned exposure times are 135.8, 135.3, and 108.1~ks for MOS1, MOS2, and PN, respectively.
For ZWCL1856, the total cleaned \textit{Chandra}/ACIS-I exposure time is 209.1~ks.

\begin{table*}[t]
\centering
\caption{X-Ray Observational Details}
\label{tab:xray_obs}
\begin{tabular}{l c l cc c c}
\hline\hline
Telescope & ObsID & Instrument & \multicolumn{2}{c}{Pointing Coordinates} & Start Date & Valid Exp. \\
\cline{4-5}
 &  &  & R.A. & Decl. &  & (ks) \\
 &  &  & (hms) & (dms) &  &  \\
\hline
\multirow{9}{*}{XMM-{\it Newton}}
& \multirow{3}{*}{0650383401}
& EPIC-MOS1 &  & & \multirow{3}{*}{2010 May 16} & 12.0 \\
&  & EPIC-MOS2 & 17 52 01.45 &  +44 40 45.8 &  & 11.9 \\
&  & EPIC-PN   &  &  &  & 5.5 \\
& \multirow{3}{*}{0903130101}
& EPIC-MOS1 & &  & \multirow{3}{*}{2022 Jul 16} & 61.4 \\
&  & EPIC-MOS2 & 17 52 01.49 & +44 40 45.8 &  & 60.5 \\
&  & EPIC-PN   &  &  &  & 52.4 \\
& \multirow{3}{*}{0903130201}
& EPIC-MOS1 &  & & \multirow{3}{*}{2022 Jul 18} & 62.4 \\
&  & EPIC-MOS2 & 17 52 01.49 & +44 40 45.8 &  & 62.9 \\
&  & EPIC-PN   &  &  &  & 50.2 \\
\hline
\multirow{11}{*}{{\it Chandra}}
 & 19752 & ACIS-I & 18 56 54.20 & +66 20 06.00 & 2018 Jun 03 & 36.8 \\
 & 26933 & ACIS-I & 18 56 36.00 & +66 22 48.00 & 2024 Jul 31 & 22.9 \\
 & 27204 & ACIS-I & 18 56 36.00 & +66 22 48.00 & 2024 Aug 28 & 24.5 \\
 & 27205 & ACIS-I & 18 56 36.00 & +66 22 48.00 & 2024 Aug 01 & 33.9 \\
 & 27206 & ACIS-I & 18 56 36.00 & +66 22 48.00 & 2024 Jan 20 & 11.8 \\
 & 27207 & ACIS-I & 18 56 36.00 & +66 22 48.00 & 2024 Feb 26 & 14.9 \\
 & 27208 & ACIS-I & 18 56 36.00 & +66 22 48.00 & 2024 Mar 02 & 10.7 \\
 & 27209 & ACIS-I & 18 56 36.00 & +66 22 48.00 & 2024 Jul 31 & 21.2 \\
 & 29201 & ACIS-I & 18 56 36.00 & +66 22 48.00 & 2024 Jan 20 & 10.0 \\
 & 29293 & ACIS-I & 18 56 36.00 & +66 22 48.00 & 2024 Feb 26 & 10.8 \\
 & 29294 & ACIS-I & 18 56 36.00 & +66 22 48.00 & 2024 Mar 02 & 11.6 \\
\hline
\end{tabular}
\end{table*}

\section{Search for inverse Compton emission in MACSJ1752} \label{app:IC}

Although the broken power-law model provides an adequate description of the X-ray SB profile along the NE relic in MACSJ1752, we find residual excess emission near the relic location. 
One possible origin of this excess is nonthermal IC emission from the same relativistic electron population responsible for the radio relic \citep[e.g.,][]{Million2009, Wik2009, Wik2011, Botteon2016b}. 
To phenomenologically account for this excess in the SB profile, we added a Gaussian component to the broken power-law model. 
This marginally improves the fit, lowering the reduced chi-square to $\chi^2_{\rm red}=1.0$ (d.o.f. = 26), while the other best-fit parameters remain consistent with the single broken-power-law model within their uncertainties.
Since the radial range of the Gaussian component overlaps with masked point sources, we repeated the analysis with varying point-source mask sizes and found the inferred excess to remain consistent.

However, this excess emission at $E\lesssim10$~keV alone does not constitute a robust detection of IC emission because of thermal X-ray photons can contribute substantially in this energy range. 
Confirmation would require a statistically significant nonthermal X-ray component whose spatial and spectral properties are consistent with the radio-emitting electron population, after accounting for the thermal ICM emission and background systematics \citep[e.g.,][]{Finoguenov2010, Ota2014, Wik2014}.
Therefore, we regard the possible IC contribution as tentative, and deeper X-ray observations, ideally with hard-X-ray coverage from facilities such as {\it NuSTAR} (ObsID: 71201005 -- PI: T\"{u}mer), are required to distinguish IC emission from thermal or background-related origins.

\section{Systematic uncertainty estimation on the Mach number} \label{app:sys}

Because the measured temperature can depend on several modeling assumptions, we quantify the systematic uncertainty on the shock temperature jump.
We repeated the spectral fitting for a set of systematic variants and compared the resulting post-shock-to-pre-shock temperature ratio with that from the fiducial model. 
For each case, the temperature jump was calculated as $R=kT_{\rm post}/kT_{\rm pre}$, and the systematic shift was defined as $\Delta R=R-R_0$, where $R_0$ is the baseline value.
The tested systematic categories include the Galactic hydrogen column density, ICM abundance, sky-background normalization, and non-X-ray-background normalization. 
For each category, we adopted the largest absolute shift among the corresponding low and high variations as the systematic contribution. 
The total systematic uncertainty was then obtained by adding the individual category contributions in quadrature. 
The resulting systematic uncertainty of the temperature jump Mach number is $0.2$.

\section{Excess surface brightness emission along the SW relic of MACSJ1752} \label{app:excess}

\begin{figure*}
    \centering
    \includegraphics[width=1\linewidth]{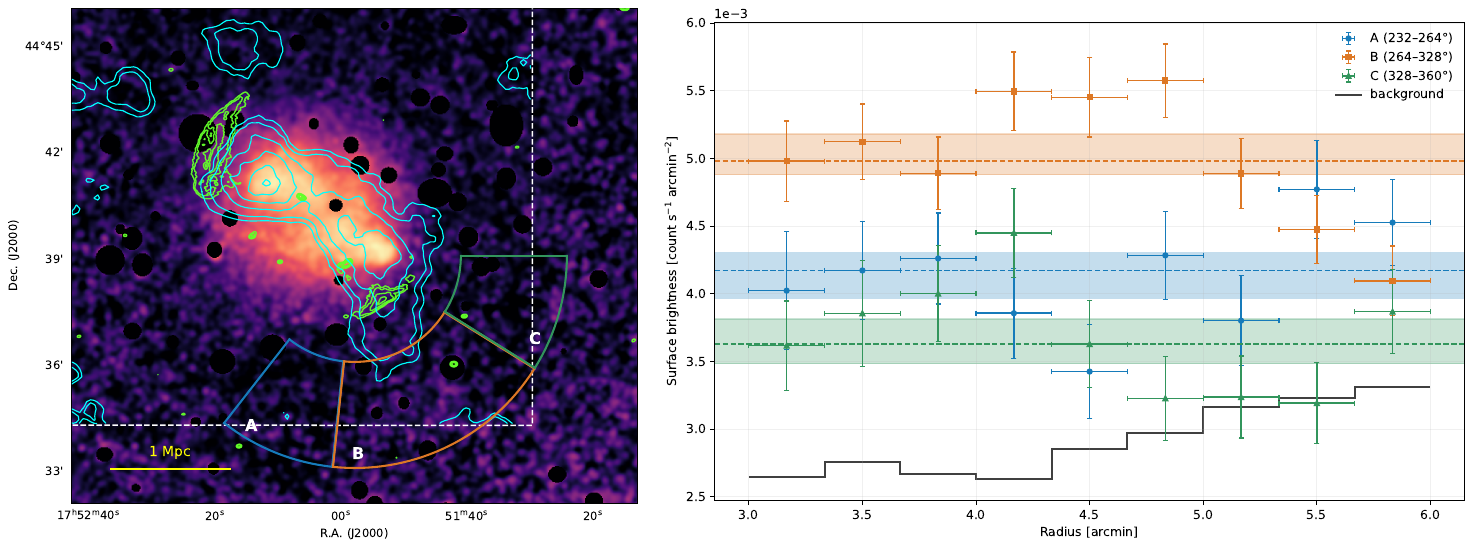}
    \caption{X-ray excess emission along the SW relic in MACSJ1752.
             Left: Exposure-corrected and smoothed XMM--Newton X-ray image, displayed with a logarithmic color scale. 
             Point sources are masked in black. 
             Cyan contours show the signal-to-noise map at $2,3,5,7,9,$ and $11\sigma$.
             The colored panda sectors used for the extraction are A (blue), B (orange), and C (green), while the white dashed line marks the mass-map footprint. 
             Right: X-ray SB profiles for sectors A, B, and C. 
             Points show radial SB measurements with $1\sigma$ uncertainties from counting statistics and the shared sky-background error. 
             Shaded bands show the central 68\% intervals of the Monte Carlo distributions of the sector medians.
             Black solid line indicates the background level along the sector B.
             The B sector exhibits an X-ray SB excess relative to sectors A and C, coincident with an extension in the weak-lensing mass map.
             }
    \label{fig:macsj1752_putative_fil}
\end{figure*}

The SB profile along the SW relic of MACSJ1752 is relatively flat at large radii (Figure~\ref{fig:macsj1752_profile}).
The weak-lensing mass distribution (cyan contours in Figure~\ref{fig:macsj1752_putative_fil}) also shows an elongation in the same direction.
To investigate whether the enhanced X-ray emission may be associated with a putative filamentary structure, we compare the SB profile along the SW relic with those extracted from two adjacent regions.

Figure~\ref{fig:macsj1752_putative_fil} shows the extraction regions and their corresponding SB profiles.
The profile along the SW relic direction (region B) lies systematically above those of the two adjacent regions over much of the radial range, with a median SB exceeding those of regions A and C by $\sim$19\% and $\sim$37\%, respectively.
We estimate the uncertainties in these median differences using $10^6$ paired bootstrap resamples, resampling the same radial-bin indices across all three sectors to preserve their cross-sector correlations.
The B--A and B--C differences correspond to $2.5\sigma$ and $4.6\sigma$, respectively.
This enhancement is spatially aligned with the elongation seen in the weak-lensing mass distribution, hinting at a possible large-scale filamentary structure.

\section{Complementary temperature maps} \label{app:temp}

\begin{figure}
    \includegraphics[width=1\linewidth]{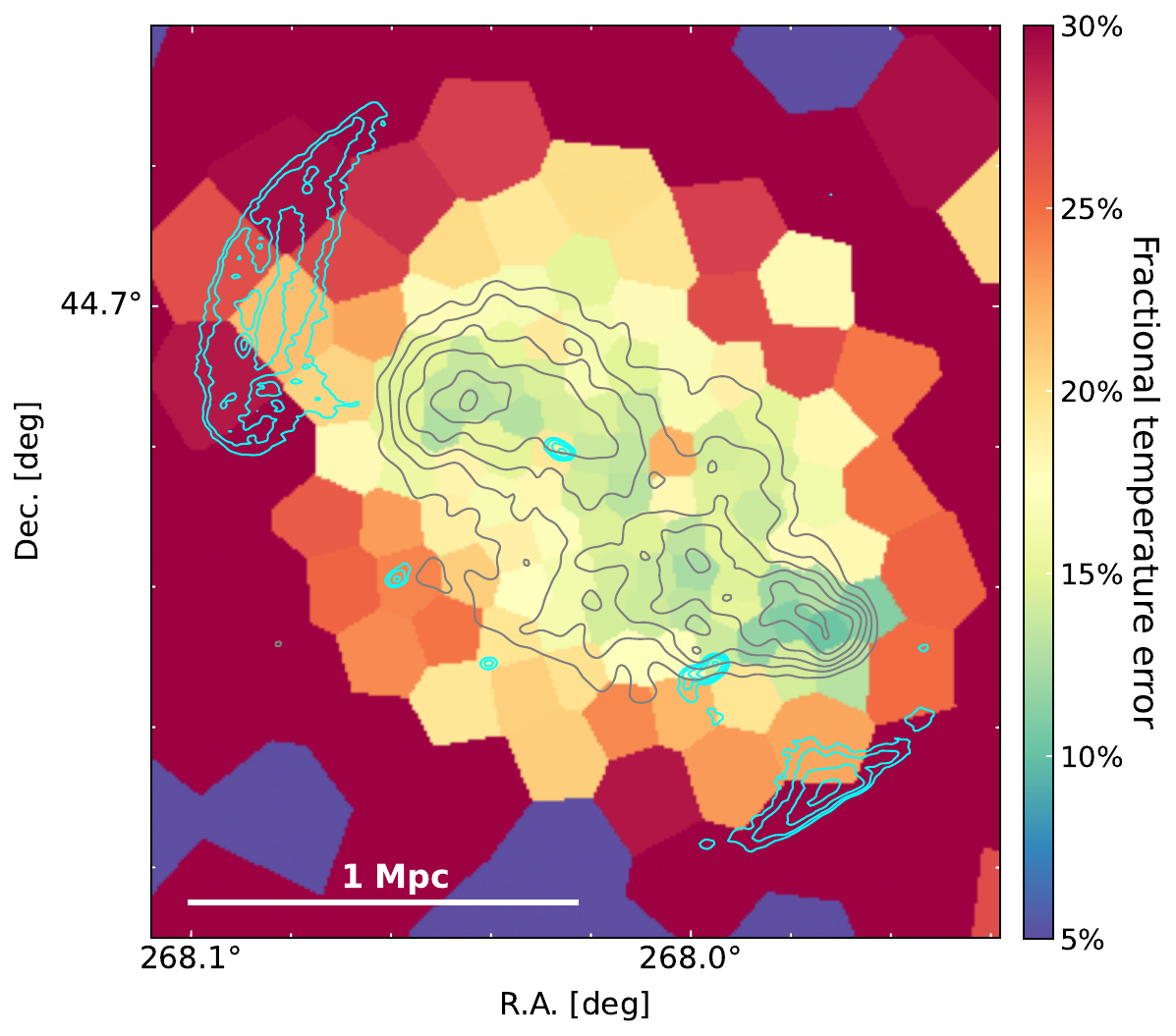}
    \caption{Fractional temperature uncertainty, corresponding to the temperature map shown in Figure~\ref{fig:macsj1752_thermo}.
             The central cluster region is typically constrained at the $\lesssim20\%$ level, with larger uncertainties toward the outskirts.
             Bins for which the temperature could not be constrained are set to zero.
            }
    \label{fig:macsj1752_temp_err}
\end{figure}

Figure~\ref{fig:macsj1752_temp_err} shows the fractional uncertainty of the temperature map of MACSJ1752.
This map provides a measure of the spatial reliability of the temperature structures discussed in Section~\ref{subsec:thermo}.
The temperatures are generally constrained to within $\lesssim20\%$ in the central cluster region, while the uncertainties increase toward the outskirts.

\begin{figure*}
    \includegraphics[width=1\linewidth]{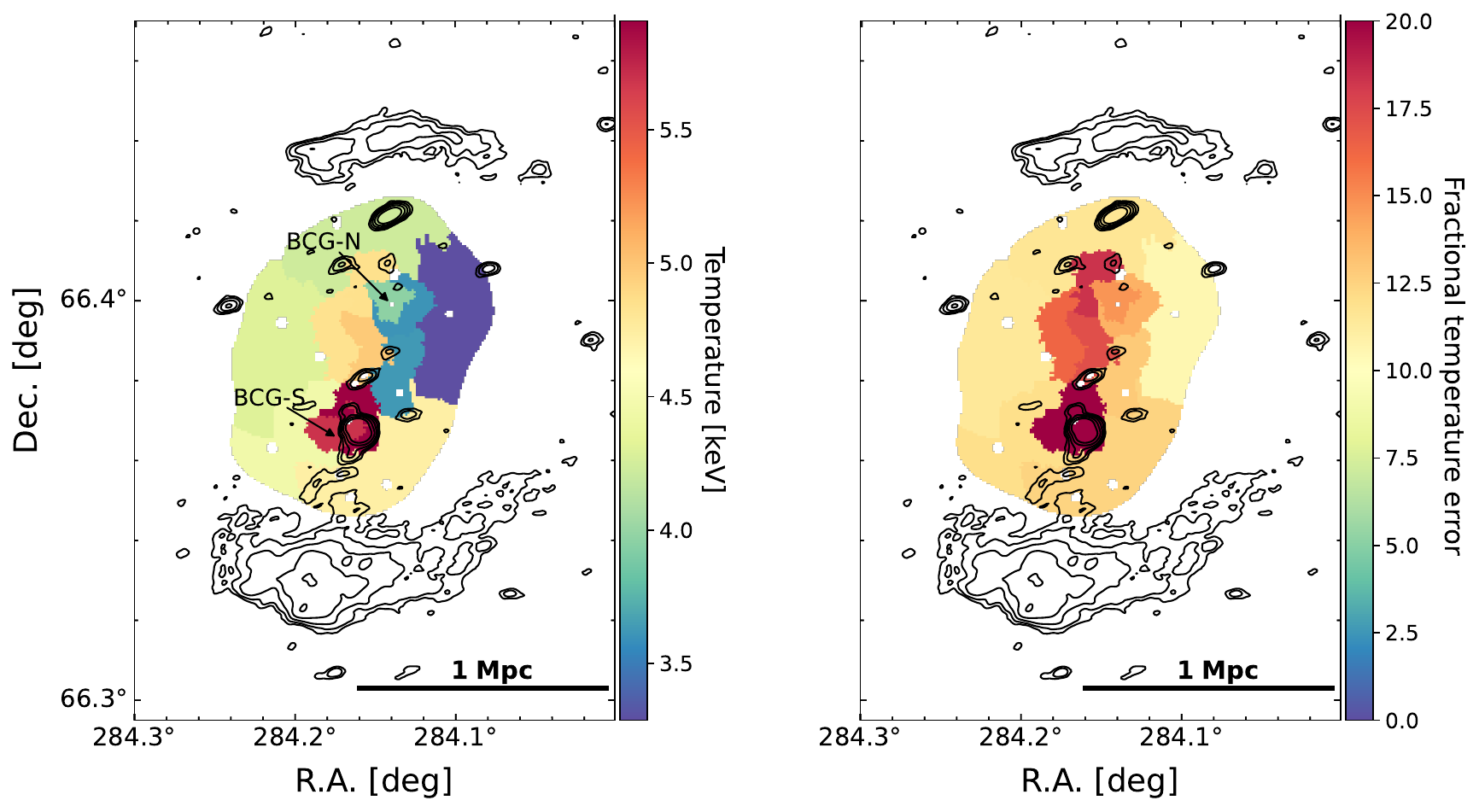}
    \caption{Same as Figure~\ref{fig:zwcl_temp_map}, but constructed using spatial bins with a target signal-to-noise ratio of 20.
             The lower signal-to-noise threshold provides finer spatial sampling than the fiducial map, enabling a closer examination of small-scale temperature structure.
             The two temperature maps are broadly consistent within $2\sigma$.
            }
    \label{fig:zwcl1856_temp_sn20}
\end{figure*}

Figure~\ref{fig:zwcl1856_temp_sn20} shows ZWCL1856's temperature map constructed using spatial bins with a target signal-to-noise ratio of 20.
Compared with the fiducial map constructed at a target signal-to-noise ratio of 30, this provides higher spatial resolution at the expense of larger statistical uncertainties, allowing us to examine smaller-scale temperature structures.
The temperature maps constructed with target signal-to-noise ratios of 30 and 20 are generally consistent within the $2\sigma$ level, indicating that the main temperature structures are robust against the choice of spatial binning.

\section{Western surface brightness profile in MACSJ1752} \label{app:macsj1752_west}

\begin{figure*}
    \includegraphics[width=1\linewidth]{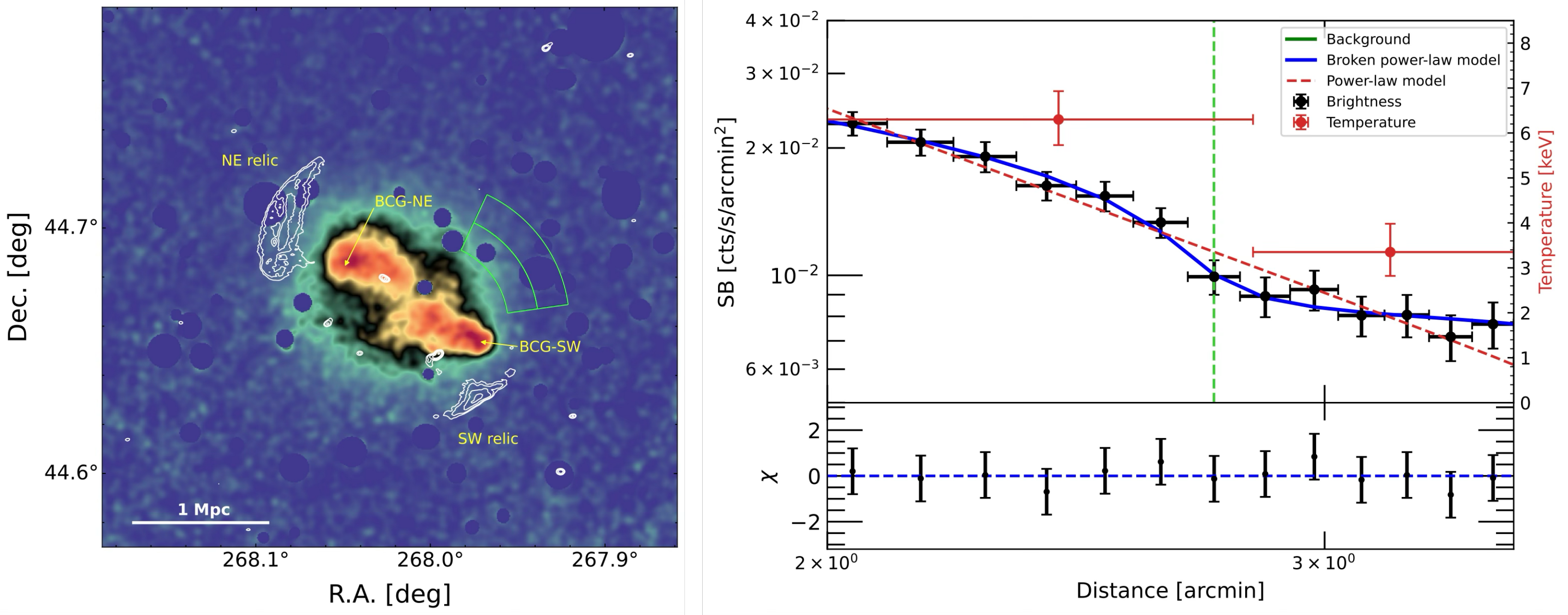}
    \caption{X-ray SB profile in the 0.5--2~keV along the putative shock direction guided by the temperature map.
             Left: Same as in Figure~\ref{fig:main} for MACSJ1752 where the point sources are masked.
             Right: The X-ray SB profile is extracted in the western shock region.
             The best-fit broken power-law model ($\chi^2_{red}=0.3$; d.o.f. = 8) is shown in blue while the best-fit single power-law model ($\chi^2_{red}=0.9$; d.o.f. = 10) is shown as a red dashed line.
             The background level is approximately $2\times10^{-3}$ and lies outside the plotted range.
             The horizontal error bars indicate the bin widths.
             The best-fit breaking radius is marked by a vertical green dashed line.
             The temperature measurements in two bins are marked in red.
             Despite the temperature drop, the SB profile does not provide compelling X-ray evidence for a shock in this region.
            }
    \label{fig:macsj1752_west_sb}
\end{figure*}

Figure~\ref{fig:macsj1752_west_sb} shows the SB and temperature profiles along the direction of the putative shock. 
The power-law ($\chi^2_{red}=0.9$; d.o.f. = 10) and broken-power-law ($\chi^2_{red}=0.3$; d.o.f. = 8) fits do not provide compelling evidence for an SB discontinuity, despite the observed factor-of-two temperature drop ($T_{post}=6.3^{+0.6}_{-0.6}$ and $T_{pre}=3.4^{+0.6}_{-0.5}$).
The low $\chi^2_{red}$ of the broken-power-law fit may reflect conservative SB uncertainties and the flexibility of the model given the small number of degrees of freedom, and we therefore do not interpret the improved fit statistic as strong evidence for a discontinuity.
In addition, the broken-power-law fit yields a density compression factor of $C=4.9\pm2.3$. 
The best-fit value formally exceeds the maximum compression factor of $C=4$ permitted for a single-fluid adiabatic shock with $\gamma=5/3$. 
The large fitted value may reflect statistical uncertainty, projection effects, geometric assumptions, or degeneracy with the density slopes. 
Given the lack of a statistically significant SB discontinuity and the formally unphysical compression factor, the feature is unlikely to represent an X-ray shock.

\bibliography{xray}{}
\bibliographystyle{aasjournal}

\end{document}